\documentclass[preprint]{aastex}
\pdfoutput=1
\newcommand{\p}{$\pm$}

\newcommand{\beam}{beam$^{-1}$}

\newcommand{\edot}{\.{E}}
\newcommand{\msun}{M$_{\sun}$}

\newcommand{\g}{G54.1+0.3}
\newcommand{\pdot}{\.{P}}

\shorttitle{Radio Properties of G54.1+0.3}
\shortauthors{Lang et al.}

\slugcomment{Accepted for publication in the Astrophysical Journal}

\begin{document}

\title{The Radio Properties and Magnetic Field Configuration in the Crab-like Pulsar Wind Nebula G54.1+0.3}
\author{Cornelia C. Lang}
\affil{Department of Physics \& Astronomy, 703 Van Allen Hall, University of Iowa, Iowa City, IA 52242}
\email{cornelia-lang@uiowa.edu}

\author{Q. Daniel Wang}
\affil{Department of Astronomy, University of Massachusetts, Amherst, MA 01002}

\author{Fangjun Lu\altaffilmark{1}}
\affil{Key Laboratory for Particle Astrophysics, Institute of High Energy Physics, Chinese Academy of Sciences, Beijing 100049, China}
\altaffiltext{1}{Department of Astronomy, University of Massachusetts, Amherst, MA 01002}

\and
\author{Kelsey I. Clubb\altaffilmark{2}}
\affil{Department of Physics \& Astronomy, Van Allen Hall, University of Iowa, Iowa City, IA 52242}
\altaffiltext{2}{Now at: Department of Physics and Astronomy, San Francisco State University
1600 Holloway Avenue, San Francisco, CA 94132}

\begin{abstract}

We present a multifrequency radio investigation of the Crab-like pulsar wind nebula (PWN) G54.1+0.3
using the Very Large Array. The high resolution of the observations reveals that \g~has a complex radio structure which 
includes filamentary and loop-like structures that are magnetized, a diffuse extent similar to the associated diffuse X-ray emission. But the radio and X-ray structures in the central region differ strikingly, indicating that they trace very different forms of particle injection from the pulsar and/or particle acceleration in the nebula.
No spectral index gradient is detected in the radio emission across the PWN, 
whereas the X-ray emission softens outward in the nebula. 
The extensive radio polarization allows us to image in detail the
intrinsic magnetic field, which is well-ordered and reveals that 
a number of loop-like filaments are strongly magnetized. In addition, 
we determine that there are both radial and toroidal components to
the magnetic field structure of the pulsar wind nebula. 
Strong mid-IR emission detected in {\it Spitzer} Space Telescope data 
is closely correlated with the radio emission arising from the 
southern edge of \g. In particular, the distributions of radio and X-ray emission compared
with the mid-IR emission suggest that the PWN may be interacting with this interstellar
cloud. This may be the first PWN where we are directly
detecting its interplay with an interstellar cloud that has survived the impact of the
supernova explosion associated with the pulsar's progenitor.
\end{abstract}

\keywords{ISM: Individual: Alphanumeric: G54.1+0.3, ISM: Supernova Remnants}

\section{Introduction}

Crab-like supernova remnants, also known as pulsar wind nebulae (PWNe), 
provide unique laboratories for studying the interaction between high-energy pulsar wind materials 
and supernova ejecta. The synchrotron emission (normally observed 
at radio and X-ray wavelengths) arises from ultra-relativistic particles 
in reverse-shocked pulsar wind materials (Rees \& Gunn 1974; Reynolds \& Chevalier 1984). 
SNR G54.1+0.3 (or hereafter, G54.1+0.3) is considered to be a Crab-like source and we
will refer to it and other similar objects as PWNe. 
Recent high-resolution {\it Chandra} observations (1\arcsec~or better) 
have revealed physical features similar to the Crab: 
a bright X-ray ring surrounding a central point source (corresponding to a 136 ms pulsar; Camilo et al. 2002), 
a western jet perpendicular to the ring, and an eastern protrusion, all
 in addition to the known low-brightness, diffuse component (Lu et al. 2002).
The X-ray spectra of the above features are all nonthermal and steepen
with increasing distance from the central pulsar, indicating significant particle energy evolution. 
The distance to G54.1+0.3 has been recently revisited by an HI absorption
and $^{13}$CO emission study (Leahy et al. 2008), which suggests that the distance to the PWN is 
approximately d=6.2 kpc, based on the morphological association with a molecular
cloud. This is slightly farther than the previous distance, estimated to be 5 kpc (Lu et al. 2002). 
Here, we adopt the 6.2 kpc distance, which implies that 1\arcmin=1.8 pc. 

However, unlike the Crab, the diffuse X-ray emission in \g~extends well
beyond the central X-ray torus/jet structures. This extensive, low surface
brightness X-ray component corresponds well to the radio emission associated with 
this object. Reich et al. (1985) first identified \g~as a small-diameter (1.5\arcmin)
radio source and suggested its nonthermal nature at 4.8 GHz. Velusamy \& Becker (1988) 
present Very Large Array (VLA) of the National Radio Astronomy Observatory 
(NRAO)\footnotemark\footnotetext{The NRAO is a facility of the National Science Foundation 
operated under cooperative agreement by Associated Universities, Inc.} observations 
of \g~at 4.8 and 1.4 GHz with resolutions of 5\arcsec~and 14\arcsec, respectively. 
They confirm the nonthermal nature of the radio source, its center-filled morphology and 
linear polarization ($\sim$10\%). The similarity of the extent of diffuse X-ray and
radio emission also differs from that of the Crab, where the radio emission is much
more extensive than the X-ray component. Similarities in the size of radio and X-ray 
PWNe have been observed in sources such as G292.0+1.8, the PWN powered by B1509-58, and 3C58 
(Gaensler \& Wallace 2003; Gaensler et al. 2002; Slane et al. 2004), 
and have interesting implications for the particle energetics and source structure.

Although detailed, multi-wavelength studies have been carried out for a growing number
of PWNe (see recent review by Gaensler \& Slane 2006), many issues relevant to Crab-like
sources remain unresolved: (1) determining which acceleration mechanism 
can produce both the radio-emitting particles in addition
the to the X-ray emitting particles is difficult (Atoyan 1999; Arons 2002; Bientenholz et al. 
2004); (2) the absence of outer shells in many of these PWNe (e.g., 3C58 and the Crab). This
question is related to the ages of the SNRs as well as the the properties of both the
SN progenitors and the surrounding interstellar medium; and (3) the role and geometry of the magnetic 
field in the inner and outer regions of PWNe are not well-studied. Radio polarization studies have been carried out for a 
few PWNe (e.g., Crab, Vela, 3C58, G292.0+1.8, Boomerang) and show that the magnetic field
structures in PWNe are highly-organized and often oriented radially toward the outer
parts of the nebula. However, the transition from the predicted torodial magnetic 
field orientation at the very center of the PWN near the location of the pulsar (Bucciantini et al. 2005), 
to the outer, diffuse parts of the PWN has not been fully studied. 

Because of its similarities (and important differences) to the Crab nebula, a multi-wavelength study of 
\g~can therefore provide insight as to many
of the above unresolved issues. Understanding the nature of the diffuse X-ray emission and its relationship to the 
diffuse radio emission in \g~is crucial for understanding the energetic history of the
PWN and its expansion into the surrounding interstellar medium. Archival VLA radio 
observations of \g~(Velusamy \& Becker 1988) have relatively short 
integration times ($\sim$45 minutes at each frequency in the C array configuration) and 
low resolution (5-14\arcsec). Therefore, we present new multi-frequency, multi-configuration 
VLA observations of \g. These observations, which have resolutions of $\sim$3-5\arcsec, 
allow us to (1) characterize the radio emission on scales more comparable to the {\it Chandra} 
X-ray observations, (2) search for large-scale diffuse emission surrounding the central 
radio source, (3) study the radio spectral index 
variations across \g, (4) investigate the distribution and strength of polarized emission
in \g, and (5) determine the Faraday rotation toward \g~and, ultimately, the intrinsic 
magnetic field structure of the PWN. The observations and data reduction are presented in $\S$2, 
the results in $\S$3, and the discussion and interpretation of results follow in $\S$4.

\section{Observations and Imaging}

VLA multifrequency observations of \g~were made during 2002-2004 using the B, C, and
D array configurations at 1.4, 4.7 and 8.5 GHz. Table 1 summarizes the observational details. At all frequencies, 
J1331+305 and J1924+334 were used as flux and phase calibrators. Standard procedures 
for calibration, editing and imaging were carried out using the Astronomical Image
Processing Software (AIPS) of the NRAO. Data from different arrays were combined to 
produce final images at each frequency. Table 2 summarizes the image parameters. In order
to measure flux densities, the images were corrected for primary beam attenuation, but
the images shown in the paper have not been corrected for this attenuation as the correction causes
the rms noise level to increase near the edges of the image, particularly at 8.5 GHz. 

Polarization calibration was done for all three frequencies using the frequent observations
of the phase calibrator J1924+334 in the B and C array observations. In the D array observations,
polarization calibration was carried out only at 4.7 and 8.2 GHz; the 1.4 GHz observations did
not have enough parallactic angle coverage for adequate polarization calibration. 
Stokes' Q and U images were made at 4.7 and 8.2 GHz and images of the polarized intensity 
(I$_p$=$\sqrt{Q^2 + U^2}$) were created. In order to study the Faraday rotation toward \g, 
additional Stokes' Q and U images were made at 4.585, 4.885, 8.085, and 8.465 GHz. 
For each of these IFs, images of the polarization angle (PA=$\case{1}{2}$Arctan($\case{U}{Q}$)) 
and the error in polarization angle were created. Using the polarization angle and error images at the four 
observed frequencies, the rotation angle as a function of wavelength was fitted to a $\lambda^{2}$ law with the AIPS
algorithm RM. This task produces images of the distribution of the rotation measure and also the distribution of
the orientation of the magnetic field intrinsic to the source by correcting the observed
position angles for Faraday rotation. 

\section{Results}

\subsection{Radio Continuum Morphology}

\subsubsection{Structure of \g}

Figures 1 and 2 show the detailed structure of the \g~radio source at 4.7 and 8.2 GHz. The radio
continuum morphology at all frequencies (including 1.4 GHz) is nearly identical. The figures show that the source
extends for approximately 2.5\arcmin~$\times$2.0\arcmin~(4.5 $\times$ 3.6 pc at d=6.2 kpc), 
with diffuse radio emission elongated in the NE-SW direction. 
The brightest emission is concentrated in the central 1-1.5\arcmin~area (1.8 to 2.7 pc), 
with intensities in this area of $\sim$1.5 mJy beam$^{-1}$ at 4.7 GHz. 
The radio emission gets considerably weaker ($\sim$0.1 mJy beam$^{-1}$ 
at 4.7 GHz) moving outwards from the center of the nebula. Velusamy \& Becker (1988) 
also note these same central features of the brightness distribution in their lower resolution observations.

The high resolution ($\sim$3\arcsec) observations presented here reveal several new features in the
morphology of \g. The brightness peaks near the center of \g, in several very pronounced 
ridges which appear to be 
tracing a ring or torus of $\sim$ 20$-$30\arcsec~diameter.  In addition, there is a central cavity in the distribution 
of radio emission near RA, DEC (J2000): 19 30 30, 18 52 15. Slices of intensity across of the source indicate that the 
intensity in the central cavity is just a fraction of that contained in the surrounding ridges: 0.7 mJy/beam 
compared to 1.3 mJy/beam at 8.5 GHz, (this trend in brightness is also observed at 4.7 GHz and 1.4 GHz). Figure 3 shows
a slice of intensity across the center of \g. Velusamy \& Becker (1988) also report on this decrease of intensity 
at the very center of \g, but did not have the resolution or sensitivity to distinguish the ring. 

Figure 1 also reveals a number of filamentary structures in the NE and S portions of the nebula (labeled in Figure 1). 
In the NE part of \g, several elongated radio structures have sizes of $\sim$30-45\arcsec~(0.7-1 pc) 
near RA, DEC (J2000): 19 30 32, 18 52 15. Two of these features 
are mainly aligned with the NE-SW orientation of the radio nebula and suggestive of a flow.  
The third feature is oriented perpendicular to the NE-SW axis.  In the southern part of the nebula, a number 
of loop-like structures are detected, ranging in size from 15-30\arcsec~(0.45-0.9 pc). Such filamentary and 
loop-like features are observed in the diffuse emission of several other PWNe, including both the Crab and 3c58, 
with similar physical sizes in all cases of 0.6 to 1 pc (Reynolds \& Aller 1988; Velusamy 1985). 
Finally, \g~shows several ``bays'' to the NE and SW sides of the central ring or torus of radio emission 
which are similar to the optical bays in the Crab and 3C58 and the radio structure of PWN G21.5-0.9 
(Furst et al. 1988; Velusamy 1985). These are discussed further in $\S$4.6.

\subsubsection{Diffuse Emission Surrounding \g}

Since \g~lies in the Galactic plane, it is surrounded by diffuse large-scale radio emission. 
These larger structures are apparent in single-dish radio surveys of the Galactic plane 
(Caswell et al. 1985; Velusamy et al. 1986) but with resolutions of 1-3\arcmin, 
such images can not distinguish compact and shell-like features as in 
interferometric studies. Similar to the Crab, no 
shell surrounding \g~has ever been detected. Figure 4 shows the region surrounding \g~at 1.4 GHz; 
several shell-like radio sources and other large filamentary features are apparent in addition
to the PWN which lies at the center of the image. 
In particular, a new radio shell-like structure has been detected (labeled in red in the figure) 
that appears to be centered around \g, with a diameter of $\sim$8\arcmin~(14.4 pc at 6.2 kpc). The radio shell
is brightest on three sides of \g, with the faintest emission located in the SW portion of the shell. 
If the radio emission is nonthermal then this shell may be interpreted as the supernova remnant 
from the the \g~supernova explosion. However, determining the spectrum of the radio emission using
the current interferometric data is difficult. 
At 4.7 GHz (image not shown; a larger field of view than Figures 1 and 2), extended emission is 
present to the NW and SE at approximately 8\arcmin~in radius
from \g.  At 1.4 GHz, there is considerably more diffuse emission in the field of
view than at 4.7 GHz, presumably because of the better sensitivity to extended features.
In fact, the largest angular size detectable with the VLA at 4.7 GHz is only 300\arcsec~(5\arcmin),
so emission associated with this large ($\sim$8\arcmin) shell will not be observed with the VLA at
this frequency. The brightest of the extended emission has an intensity of 100-200 $\mu$Jy \beam, which 
corresponds to a signal-to-noise of $\sim$10 or less at best, and is not well represented by the {\it (u,v)} coverage 
of the 4.7 GHz observation. Therefore we are unable to provide any reliable measure of the spectral index of the 
shell-like source surrounding \g. This shell is visible partly in the VLA Galactic Plane Survey image 
(D-array survey, 1\arcmin~resolution) of Leahy et al. (2008), but it is on
a much smaller scale than their large image (field of view $\sim$1 \arcdeg). The nature of this
shell is further discussed in $\S$4.6. 

\subsection{Spectral Index of \g}

Since we have detected \g~at three frequencies (8.2, 4.7 and 1.4 GHz), 
it is possible to derive the spectral index of the radio emission. The radio emission
can be integrated over the entire source (a region of $\sim$2\arcmin~$\times$ 1.5\arcmin). 
In all cases, the integrated flux density measurements have come from identical regions of \g. 
However, \g~lies in a crowded part of the Galactic plane (see above), many large scale extended
features are present in 1.4 GHz image, which has a much larger field of view (30\arcmin) compared
with the 4.7 or 8.2 GHz images (FOV $\sim$ 5-9\arcmin). The background flux levels in the 1.4 GHz
image are much higher and variable over the image size than at other frequencies, and these
variations contaminate the flux density measurements at 1.4 GHz. Therefore, we have corrected
the 1.4 GHz measurement to yield S$_{1.4}$=433.0\p30.0 mJy, S$_{4.7}$=327.0\p25 mJy, 
and S$_{8.5}$=252.0 \p20.0 mJy. The average value of the integrated spectral index for 
\g~is $\alpha$=$-$0.3\p0.1. The spectral index between 1.4 GHz
and 4.7 GHz has a value of $\alpha$=$-$0.2\p0.1 and between 4.7 and 8.5 GHz, $\alpha$=$-$0.5\p0.2. 
In addition, Figure 5 shows the spectrum of \g~based on the integrated flux density values and the best fit
to this plot gives a spectral index value of $\alpha$=$-$0.28.  

In order to search for variations of the spectral index across the source, the spectral index can
be calculated pixel by pixel across \g~using the AIPS task COMB (opcode=SPIX). 
Figure 6 shows the distribution of spectral index 
between 1.4 and 4.7 GHz in greyscale and overlaid by contours of 4.7 GHz continuum emission.
This image was created by using data from matched arrays so that the spatial frequency sampling (i.e., {\it (u,v)} coverage) is consistent between both frequencies. 
In practice, this means that the 1.4 GHz image is constructed from B-array data only, 
the 4.7 GHz from C-array data only and the 8.2 GHz from D-array data only. 
Further, the spectral index images were obtained by convolving both images to the same resolution (5\arcsec) and 
determining the spectral index value at every pixel. The spectral index between 4.7 and 1.4 GHz in \g~does not show any
variation across the nebula, with a fairly constant value of $\alpha$=$-$0.2. The spectral index between 4.7 and 8.2 GHz
also does not show any significant variation across \g, but has a slightly steeper average value, $\alpha$=$-$0.3. 

\subsection{Polarization}

There is strong linear polarization detected in \g~at both 4.7 and 8.5 GHz. 
The polarization at 1.4 GHz was significantly weaker ($\sim$10\% 
fractional polarization) presumably because of bandwidth depolarization
across the 50 MHz bandwidth at 1.4 GHz. Figure 7 
shows polarized intensity at both 4.7 and 8.5 GHz. These figures illustrate that there
is significant polarized intensity arising from \g~that closely follows the total 
intensity structure. The polarization is concentrated along the two central radio ridges, 
and the polarized intensity peak is offset by $\sim$10-15\arcsec~from the center of the PWN. 
Diffuse polarized intensity extends over much 
of the nebula. Figure 7 (top) shows that several of the 
radio loop-like features along the NW-SE extensions of the nebula are polarized.  There
are two striking linear features that extend for up to 30\arcsec~(0.9 pc) in extent; 
they are labeled in Figure 7 (top). 
The fractional polarization was sampled across the nebula and varies between 20$-$50\% at 
8.2 and 4.7 GHz. Typical values of the fractional polarization are 27\% at 8.2 GHz, 23\% at 4.7 GHz. 
The highest values of fractional polarization are labeled in Figure 7 (bottom) and have values
close to 40\% at 8.5 GHz near RA, DEC (J2000): 19 30 30, 18 52 30.0 and 50\% at 8.5 GHz near RA, DEC (J2000): 
19 30 31, 18 52 50.0. 

\subsection{Rotation Measure and Intrinsic Magnetic Field Orientation}

Figure 8 shows the distribution of rotation measure (RM) in rad m$^{-2}$ toward \g, based on 
fitting the polarization angle as a function of wavelength squared (see $\S$2) for interstellar 
Faraday rotation. Across much of the \g~nebula, the values for RM are in the range of 
400-650 rad m$^{-2}$. In two regions (in the NE near RA, DEC (J2000): 19 30 33, 18 52 45, and in the SW near RA, 
DEC (J2000): 19 30 28, 18 52 15) the RM values appear to be lower by several hundred rad m$^{-2}$.
Figure 9 shows a sample of the fits of polarization angle as a function of wavelength 
squared, where each panel represents the data averaged over a few pixels from two small regions 
of the source. The errors in the RM fits are on the order of 25-75 rad m$^{-2}$. 

The vectors shown in Figure 10 represent the orientation of the magnetic field intrinsic to \g, obtained 
by correcting the observed polarization angles for the calculated RM. The magnetic field is very well 
ordered over the majority of the region which is polarized in \g~(see Figure 9, left). In particular, 
the magnetic field appears to follow 
closely the curvature of the central radio ``ring'' region. At the edges of the nebula, the magnetic field
appears to be oriented along the linear extensions which are present in both total and polarized
intensity and have a radial orientation. In the SE and SW portions of \g, the field becomes less-ordered
as the polarized intensity fades. 

\section{Discussion}
\subsection{Comparison of Radio Emission and X-Ray Emission}

Figure 11 shows a comparison of the 4.7 GHz radio emission in contours 
and the {\it Chandra} X-ray emission in colorscale (from Lu et al. 2002). 
The left panel has a colorscale stretch to emphasize the extended X-ray
emission, and the right panel is set to emphasize the concentrated central
X-ray emission. This figure, especially the left panel, illustrates that
the distribution of large-scale and diffuse X-ray emission in \g~is strikingly similar to 
the extent of the radio emission. Figure 12 shows radial profiles of brightness 
taken along the NE (left) and SW (right) portions of \g. The dotted line illustrates 
the ``edge'' of the PWN in radio emission, corresponding
approximately to the lowest radio contour level in Figure 11 (right panel). Although 
the X-ray emission drops more significantly than the radio emission, there is still
diffuse and low-level X-ray emission arising from the same volume as the radio emission
out to the lowest radio contour level. 

However, in the center of the PWN, the radio and X-ray emission are not as
well-correlated. The X-ray emission peaks at RA, DEC (J2000)= 19 30 30.1, 18 52 14.10, whereas
the maximum in radio emission occurs at RA, DEC (J2000)= 19 30 30.5, 18 52 11.7.
These positions are $\sim$6\arcsec~apart. Figure 11 (top) shows that 
the peak in X-ray emission lies in a slight depression of
radio emission, but with ridges of radio emission on either side. 
In addition, the projected orientation of the extended radio emission differs from
the orientation of jet-like X-ray features (e.g., Figure 4 from
Lu et al. 2002). These compact X-ray features in the center of \g~are oriented at angles 
of $\sim$45\arcdeg~with respect to the extended radio nebula (which runs along a NE to SW line). 
Apparently, the X-ray and radio emissions trace two rather distinct components of relativistic 
particles in the inner region of the PWN. While both emissions are apparently due to synchrotron 
radiation, their corresponding particle energies are very different ($\sim$ 100 TeV 
vs. $\sim 1$ GeV) and clearly represent different forms of pulsar injections and/or accelerations in the region.  

Several PWN have been studied recently at high-resolution in the X-ray and radio: 
G0.9+0.1 (Dubner et al. 2008), 3C58 (Slane et al. 2004), G292.0+1.8 (Gaensler \& Wallace 2003), 
and G21.5-0.9 (Furst et al. 1988; Bietenholz \& Bartel 2008). 
\g~bears resemblance to several of these PWN, but is most similar to 3C58 in the following ways: 
(1) the substructure in \g~is filamentary and magnetized and (2) the extents of the diffuse radio
and X-ray emission are comparable. Both 3C58 and \g~show evidence for filamentary and loop-like structures in
X-ray and radio emission on size scales of 0.7-1 pc (e.g., in 3C 58 these structures
are on size scales of 10-15\arcsec~for a distance of $\sim$3 kpc and those scale 
directly to loop-like features of size 20-30\arcsec~in \g~at a distance of $\sim$6 kpc). 
Slane et al. (2004) propose that the filamentary loops in 3C58 are magnetic loops torn from 
the toroidal field (associated with the pulsar) by kink instabilities. 
Figures 7 and 10 show that in \g~these filamentary loops are strongly polarized and have magnetic 
field orientations that align along their lengths. In fact, the curvature of magnetic 
structures that extend for as much as 15$-$30\arcsec~are one of the more remarkable features of \g.

The overall diffuse structure of the radio and X-ray emission 
(that extends along the NE to SW line) in \g~closely resembles that of 3C58, in which the X-ray
and radio emission fill essentially the same volume. Slane et al. (2004) report
that the orientation of the elongation could be related to the toroidal magnetic field 
of the pulsar and that the elongation should line up with the orientation of the projected
pulsar spin axis, as is roughly the case in 3C58. However, as described earlier there is a 
discontinuity between the orientation of the X-ray jet outflows and the extended radio emission in \g. 
Assuming that the pulsar spin axis is aligned with the X-ray jet-like feature, 
Ng and Romani (2004) estimate that the pulsar spin axis to be 
$\sim$90\arcdeg~with respect to N, or $\sim$45\arcdeg~with respect to the diffuse radio emission. 

\subsection{Source Energetics and Nebular Magnetic Field Strength} 

Figure 13 shows the spectrum of \g~over the range of $\sim$10$^8$ to $\sim$10$^{18}$ Hz. 
The flux in the diffuse X-ray emission is 5.2 $\times$ 10$^{12}$ ergs cm$^{-2}$ s$^{-1}$,
which corresponds to 1.73 $\mu$Jy. 
The radio spectrum (as discussed in $\S$3.2 and shown in Figure 5) is relatively flat, with
$\alpha$$_{R}$=$-$0.28, corresponding to a photon index ($\Gamma$)=$-$1.28. 
The diffuse X-ray nebula (not including the pulsar emission) 
has $\Gamma$=$-$1.96\p0.08, or $\alpha$$_{X}$ $\sim$$-$0.96. A spectrum
was made and fit for this emission, defined by the ``outer nebula'' region 
from Lu et al. (2002). There is a trend in the diffuse X-ray emission, though,
to soften outwards in the nebula: the inner X-ray ring has $\Gamma$=$-$1.6,
and the photon index gradually decreases outwards in the nebula to nearly $-$2. 
This indicates that there is cooling in the X-ray particles as they extend 
to fill a similar volume as the radio particles. 
However, the X-ray index of uncooled particles, $\Gamma$=$-$1.6, is significantly steeper 
than the radio index. This shows that the responsible particles cannot be described 
by a single power law energy distribution. 

According to Gaensler et al. (2002), the X-ray spectrum in PWNe are typically 
steeper than at radio wavelengths in part due to synchrotron losses which 
generate a break in the spectrum at the ``break frequency''. 
In this case, one can use Figure 13 to derive the break frequency. 
Figure 13 illustrates that the break frequency 
is the position where the two solid curves meet and occurs near $\nu_{b}$=50 GHz
for the X-ray spectrum of $-$0.96. However, the errors on the 
X-ray spectral fit give a large range of possible break frequencies, 
from $\nu_{b}$=3 to 200 GHz. A variety of PWNe have been shown to have a steep spectral break 
at relatively low frequencies ($\nu_{break}$ $<$ 100 GHz; Woltjer et al. 1997); examples include
3C58 (Slane et al. 2004), G292.0+1.8 (Gaensler \& Wallace 2003), 
and G21.5-0.9 (Bock et al. 2001) and \g~may be consistent with this class of 
PWNe. Here, we assume a break frequency from our radio and X-ray 
observations $\nu_{b}$=50 GHz. 

With a radio flux density of 327 mJy at 4.7 GHz, the $\alpha$=$-$0.28 spectral index 
and a distance of 6.2 kpc, the radio luminosity over the range 
of 10$^7$ Hz to 10$^{11}$ Hz is 8.7 $\times$10$^{32}$ erg s$^{-1}$.
Over a similar range in frequency, the Crab has a radio luminosity of 1.8 $\times$ 10$^{35}$ 
erg s$^{-1}$. The value for radio luminosity of \g~is several orders of magnitude below
what has been found for many other PWNe, such as the Crab. What may be
a more relevant quantity to compare is the ratio of radio luminosity to the 
spin-down power of the pulsar. For \g~this ratio (L$_R$/\edot) is 8.7 x 10$^{32}$/3.0 $\times$10$^{36}$ = 0.0003. For the Crab, we calculcate almost exactly the same number: L$_R$/\edot=
1.8 x 10$^{35}$/4.6 $\times$10$^{38}$ = 0.0004.

The magnetic field strength in the PWN may play a role in regulating its 
expansion and in the energetics of the outflowing particles. 
If the break frequency is known or estimated, and the age of the PWN is
known, then the nebular magnetic field can be calculated using the 
expression in Frail et al. (1996). The characteristic age (P/2\pdot) of the 
pulsar associated with \g~was determined to be 2900 years (Camilo et al. 
2002). Using this age and the break frequency of 50 GHz, we estimate a
magnetic field strength of $\sim$1250 $\mu$G. However, since the break frequency
is not very well known, this estimate is more uncertain.  

Alternatively, we can use the radio luminosity above and assume that there is
equipartition between the particles and the magnetic field
in the nebula. If we approximate the nebula as a sphere with diameter of 3.1 pc, then we 
derive an equipartition magnetic field strength of 38 $\mu$G.  
Another way to estimate the nebular magnetic field is by using the 
lifetime of the X-ray emitting particles. 
The particle flow velocity just down stream from the termination shock is derived by
Lu et al. (2002) to be about v=0.4 c. Assuming a simple situation in which 
the flow is radial, we estimate a lifetime of 12.5 years (e.g., radius of
PWN/the flow velocity).  
The lower limit on the mean flow velocity is unknown, but if we assume 
v=0.1 c, the lifetime of the particles will be 50 years. 
Then, assuming that the synchrotron lifetime at X-ray energies
(i.e., $\sim$1.4 keV) depends on the magnetic field strength, we can derive a magnetic
field in the nebula between 80-200 $\mu$G using the above velocities. 
Here, we understand that the magnetic field will be an upper limit as
the particles may diffuse out to the edge of the nebula instead. 
 
However, this magnetic field $-$ substantially higher than the equipartition magnetic field $-$ 
suggests that G54.1+0.3 is filled by a magnetically-dominated plasma,
consistent with the presence of the polarization (and hence the magnetic field)
organized on large scales of the nebula. Therefore,
the magnetic field strength in \g~is probably stronger than the equipartition
magnetic field, although the original wind from the pulsar is particle
dominated (Lu et al. 2002). What might be the origin of
this magnetic field? According to the convetional model for the Crab nebula 
(e.g., Kennel and Coroniti 1984), the pulsar wind is particle-dominated.
But when it is terminated, the magnetic field is expected to be amplified
in the down stream. The magnetic field can eventually become higher 
than the equipartition value. A detailed modeling, which accounts for 
the detailed magnetic field structure as reported here (see below), could 
potentially shed new light into the evolution and dynamics of the PWN.

\subsection{Rotation Measure Distribution Toward \g}

The fairly constant RM distribution across \g~indicates that we are detecting
the Faraday rotation by the interstellar medium in the direction of \g~rather
than an internal rotation process. There is a slight variation, however, of
the RM, with larger values (near 500-600 rad m$^{-2}$) being concentrated toward
the central 1\arcmin~ of the source and lower values (200-400 rad m$^{-2}$) concentrated
outwards in the nebula. In particular, there are gradients in the RM in the 
NE and SW extents of the nebula where it is thought that the PWN may be 
interacting with a component of the interstellar medium (see below).  

\subsection{Nature of the Magnetic Field Configuration in \g}

Figure 7 shows that strong linear polarization is evident across
much of \g~at both 8.5 and 4.7 GHz and that little change in polarized
intensity morpholoy or fractional polarization occurs between frequencies. 
Because the polarization extends over 
much of the PWN, it is possible to determine
the intrinsic magnetic field structure over this region. Studies that map
out the intrinsic magnetic field have only been carried out for a handful of
PWN sources (e.g., G106.6+2.9 (Kothes et al. 2006); Vela (Dotson et al. 2003); 
G21.5-0.9 (Furst et al. 1988), and G292.0+1.8 (Gaensler \& Wallace 2003)). 
A large sample is needed to fully understand the magnetic field
structure in these objects.  

Kothes et al. (2006) present an interpretation of radio polarization measurements
for G106.6+2.9, but generalized for many PWNe. They suggest that a toroidal field
will appear to be tangential to the spin axis of the pulsar, whereas the radial 
field will appear along the elongation axis of the PWN. In many cases, 
the geometry will not be straightforward because of projection effects 
and the orientation of the observer. However, 
if the spin axis of the pulsar is known, then it is possible to imagine how
the transition between the magnetic field in the inner region of the pulsar
may be projected into something observable in the outer parts of the PWN. 

In \g, the spin axis of the pulsar is assumed to be aligned roughly 
with the X-ray jets (i.e., Ng \& Romani 2004), as in the Crab nebula and 3C58. 
Therefore, a toroidal field component would appear to be oriented perpendicular to this direction,
and a radial magnetic field would align along this axis. Figure 14 shows
the orientation of the intrinsic magnetic field vectors (in the plane of the sky)
overlaid on greyscale representing the X-ray emission in \g. It is apparent
that the majority of the magnetic field vectors are aligned in the central
part of the nebula (where the X-ray emission is strongest) essentially
tangential to the spin axis, but in the outer parts of the nebula, the field is
radial and extends for the most part along the elongation. The exception is 
in the linear part of the loop-like feature in the NE part
of the PWN. There appears to be a large magnetized loop emerging from this 
part of the PWN, perhaps an expanding part of the inner field structure. The difficulty with
this is that the elongation of the diffuse nebula and the spin-axis from the X-ray are
tilted with respect to each other, so this distribution of magnetic structure is complex.  
\g~appears to exhibit a complex radial magnetic field pattern, according to the
classification of Kothes et al. (2006) by showing both radial and toroidal field components.  

Finally, the role of the magnetic field in \g~may be similar to that in 
3C 58. As described in the previous section, Slane et al. (2004) 
conclude that in 3C 58 the possible disruption of a toroidal 
magnetic field configuration as the pulsar wind expands  
is responsible for producing the radio and X-ray "loops" and 
filamentary radio polarization structures. This may be the same 
mechanism at work in \g~where a toroidal magnetic field is present
and may be disrupted in the outer parts of the nebula. 

\subsection{New Radio Shell: Large-scale Emission Around G54.1+0.3}

A clue to the nature of the new radio shell surrounding \g~can be gained by comparing the radio emission 
to the {\it Spitzer Space Telescope} 24 $\mu$m image made as part of the MIPS Legacy Project (MIPSGAL; Carey et al. 2009). 
The 24 $\mu$m emission traces stocastically-heated, small dust grains which trace massive stellar activity in the 
Galaxy. The radio continuum and 24 $\mu$m emission turns out to align very closely 
as evidenced in Figure 15. In addition to the large radio shell, Figure 15 also illustrates that
there is a significant source of 24 $\mu$m emission associated with \g~itself that has a smaller,
``loop-like'' geometry (80\arcsec~or 2.4 pc across). 
This correspondance has been noted before by Koo et al. (2008) and Slane et al. (2008). 
Figure 16 shows archival {\it Spitzer} data (ID:3647) taken at 24$\mu$m pointed 
at \g~with radio continuum contours at 4.7 GHz overlaid. However, as pointed out in $\S$3.1.2, 
the nature of the radio continuum emission is unclear and could be either free-free brehmstralung
from ionized gas or nonthermal synchrotron emission. 

Koo et al. (2008) indicate there may be some young stellar object (YSOs) sources and early B-stars 
embedded in the bright mid-infrared (mid-IR) loop that could be responsible for heating this
large-scale radio shell (i.e., ionized gas).  
However, these ionizing sources may not be strong enough to account for
the extensive radio emission in the radio shell if the total content of YSOs is
on the order of 100\msun as suggested by Koo et al. (2008). In this case, 
the correspondance might represent the coincidence of a mid-IR shell with nonthermal radio
emission. More sensitive multi-frequency radio observations which include the single dish flux 
are necessary to measure the spectral index accurately. 

\subsection{Interplay of \g~with a Mid-infrared Cloud}

As mentioned above, a more detailed view of the
correlation between \g~and the mid-IR loop is shown in Figure 16 and reveals several
morphological details: the bright radio ridges in the center of the PWN exactly
fill a depression in the mid-IR loop, and the contours of radio emission, especially at the
SW edge of the PWN very closely track the morphology of the mid-IR emission. 
In addition, an overlay between the X-ray emission in \g~and the 24
$\mu$m emission made by Slane et al. (2008) also reveals that the brightest X-ray emission
concentrated in the central region of \g~is closely correlated with the edge of
the mid-IR loop, indicating they may be interacting (see their Figure 3). 
Koo et al. (2008) have suggested that a burst of
star-formation evidenced by the presence of various near-to mid-IR sources 
(proposed to be massive YSOs), the formation of which may be triggered by the interplay
of the stellar wind from the progenitor star of PWN \g. The large 24 $\mu$m luminosity 
and the presence of massive young stellar objects (Koo et al. 2008) indicate that 
the cloud is a pre-existing interstellar cloud, instead of part of 
the SN ejecta. It would also be difficult to imagine how the ejecta
could remain so dense and cool in such a young SNR.

Several other features observed in \g~can be naturally explained with an interaction
scenario: (1) the anti-correlation between peaks of the mid-IR emission and the
distribution of RM (or similarly, polarized emission) shown in Figure 18. 
The polarization disappears abruptly at the edges of \g, before the radio continuum edge of the PWN, right at the
positions where the thermal IR emission from the cloud appears to peak. In fact, closer
inspection shows this change quite clearly: Figure 17 presents 
slices in fractional polarization at 4.7 GHz in (left) the SE part of \g, and (right) the 
SW part of \g~(see caption for exact locations of slices). In both cases, abrupt
decreases in fractional polarization are detected before the edge of the
radio continuum emission has been reached and coinciding with the peak of
the mid-IR emission. In the case of the slice along the W side of \g, the fractional
polarization falls off and then increases briefly again near the peak of the
mid-IR emission. The sudden decreases in polarization can naturally be
explained if the pulsar wind flow (hence, the magnetic field) 
changes direction because of the confinement of the mid-IR shell. 
(2) The change in direction of the pulsar wind flow will also have the
effect of changing the orientation of the intrinsic magnetic field.  Figures 10 and 14
show that the magnetic field becomes much less regular in the SW part of the PWN compared to the
well-ordered and primarily radially-oriented fields in the N of the PWN. 
(3) Overall,  the expansion of the PWN  may be confined by the cloud
which can explain the differences in the radio and X-ray emission. The
brightest X-ray features are concentrated in the center of the PWN (see
Figure 11 (top panel)  and are dominated by the jets and central X-ray
torus (e.g., Lu et al. 2002)). However, the diffuse X-ray emission and radio emission
are well-aligned and their morphology is likely to be shaped by the
ram-pressure confinement of thermalized particles. (3) The radio intensity
contours (see Figure 16) appear to be compressed SW to the pulsar, relatively to the NE, which
again can be a natural consequence of the ram-pressure confinement provided by
the IR shell to the SW. There is no clear detection of the mid-IR cloud in existing $^{13}$CO (1-0) data. 
One possibility is that the cloud could have been heated substantially and may be bright in
higher order transitions. Sensitive observations of various molecular tracers
are needed to further the study of the interplay between the \g~and the cloud.

The apparent confinement of \g~by the surrounding mid-IR cloud is somewhat
similar to that of another PWN N157B in the Large Magellanic Cloud (Wang 
et al. 2001). In the case of PWN N157B, however,
the confinement is probably due to the reflection of the supernova ejecta from
a large-scale dense cloud, in contrast to the direct interaction of \g~with
the mid-IR cloud. In both cases, the X-ray emission is largely extended,
comparable to the radio emission in size, which is apparently caused by
the bulk motion of pulsar wind materials due the direction asymmetry in the
ram-pressure confinement. The consideration of such bulk motion is
important in understanding the evolution of the relativistic particles in 
the pulsar wind of \g~and the resultant X-ray and radio emission.
Figure 19 shows a sketch of a possible interaction geometry between \g~and 
an interstellar cloud.  The morphological ``bays'' mentioned earlier in the text and illustrated
in Figure 1 are therefore likely to be more complicated that simple analogs
to the bays in the Crab or in G21.5-0.9. The nature of the bays is far
from clear and may be related to how the relativistic particles are injected
from the pulsar and/or further accelerated in the nebula. The issue may also be
complicated by the apparent confinement of the nebula by the mid-IR cloud.

In summary, this may be the first PWN where we are directly detecting
its interplay with an interstellar cloud that has survived the impact
of a supernova explosion. This interplay seems to give a natural explanation
of the morphology and polarization change of the radio emission as well as
its similarity and dissimilarity to the X-ray intensity distribution of
the \g. 

\section{Conclusions}

Here we summarize our radio observations of \g~and comparision with multi-wavelength
observations of this source and draw the following conclusions: 

(1) The higher resolution radio observations presented here reveal that 
\g~has a complex structure which closely resembles other PWNe (e.g., in 
particular, the Crab, 3C 58, and G.21.5-0.9). \g~shows wispy and filamentary
radio structures in the outer parts of the nebula that are loop-like and are 
polarized. On large scales, the radio emission in \g~ is extended along the NE$-$SW 
direction. The radio emission in the center of the nebula is concentrated into a ring or
torus-like object with a larger radius that the X-ray torus. 

(2) Comparisons between \g~in the radio and X-ray are striking as there is
much similarity at these two different frqeuencies. Both the radio and X-ray 
emission have comparable diffuse extents with the presence of filamentary
and loop-like structures as mentioned above. However,
the inner part of \g~is dominated by bright X-ray emission associated with the jet-like
features. These structures are offset from the central radio features, which peak
further outwards in the nebula than the X-ray emission. The radio emission also 
tends to dominate the nebula on larger scales.  

(3) No spectral index gradient is detected in the radio emission across the PWN, 
whereas the X-ray emission softens outward in the nebula, 
consistent with the synchrotron cooling of the X-ray-emitting particles 
in a magentic field of about 100 $\mu$G. 

(4) The strong radio polarization present in \g~allows us to image in detail the
intrinsic magnetic field across the PWN. Clear radial and toroidal components of the
magnetic field are detected in \g, which is an intermediate morphology for the field, 
according to the classification scheme of Kothes et al. (2006). 

(5) Strong mid-IR emission from a bright loop is closely correlated with the southern
edge of \g. In particular, the distribution radio and X-ray emission compared
with the mid-IR emission suggest that the PWN may be interacting with this interstellar
cloud. Figure 19 shows a sketch of such a possible interaction. 

(6) An interaction between the PWN \g~and the interstellar cloud (traced in mid-IR emission) 
may naturally explain the differences in the orientation of X-ray and radio emission in \g, the 
change of magnetic field orientation and the RM distribution and the apparent
compression of the radio emission along the southern part of \g. 
This may be the first PWN where we are directly detecting its interplay with an
interstellar cloud that has survived the impact of the supernova explosion associated
with the pulsar's progenitor. 

\acknowledgements

The authors thank an anonymous referee for helpful comments. We also thank 
Dr. Wenwu Tian for helpful discussions regarding the assocation of
\g~with surrounding interstellar media. F.J. Lu is supported by the Nature Science Foundation 
of China through grants 10533020 and by National Basic Research Program of China (973 Program 2009CB824800).

\clearpage

\begin{figure*}
\includegraphics[width=1.0\textwidth,angle=0]{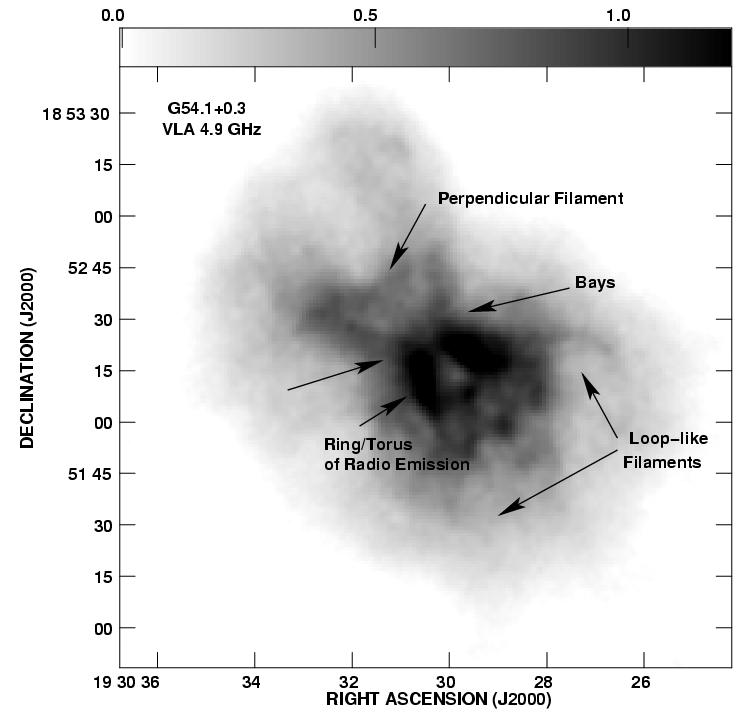}
\caption{VLA 4.7 GHz continuum image of G54.1+0.3 shown in greyscale representing 0 to 1.25 mJy beam$^{-1}$ (see axis on top of
figure for greyscale range in mJy beam$^{-1}$). The resolution of
this image is 3.28\arcsec~$\times$3.21\arcsec, PA=$-$41.6\arcdeg~and the rms noise is 15 $\mu$Jy \beam.}
\end{figure*}

\begin{figure*}
\includegraphics[width=1.0\textwidth,angle=0]{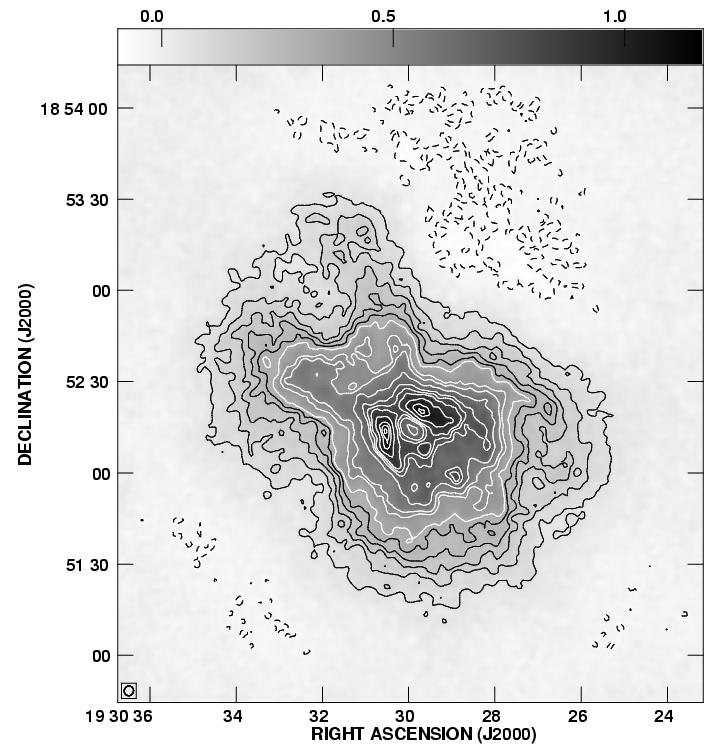}
\caption{VLA 8.2 GHz continuum image of G54.1+0.3 shown in greyscale representing 0 to 1.25 mJy beam$^{-1}$ (see axis on top of figure for greyscale range in mJy beam$^{-1}$) and contour levels representing 
-5, 5, 10, 15, 15, 20, 25, 30, 40, 45, 55, 65, 70, 75, 85, 95, 100 and 104 times the rms level of 10 $\mu$Jy \beam. The resolution of
this image is 3.22\arcsec~$\times$3.01\arcsec, PA=$-$20.0\arcdeg.}
\end{figure*}

\begin{figure}
\includegraphics[width=1.0\textwidth,angle=270]{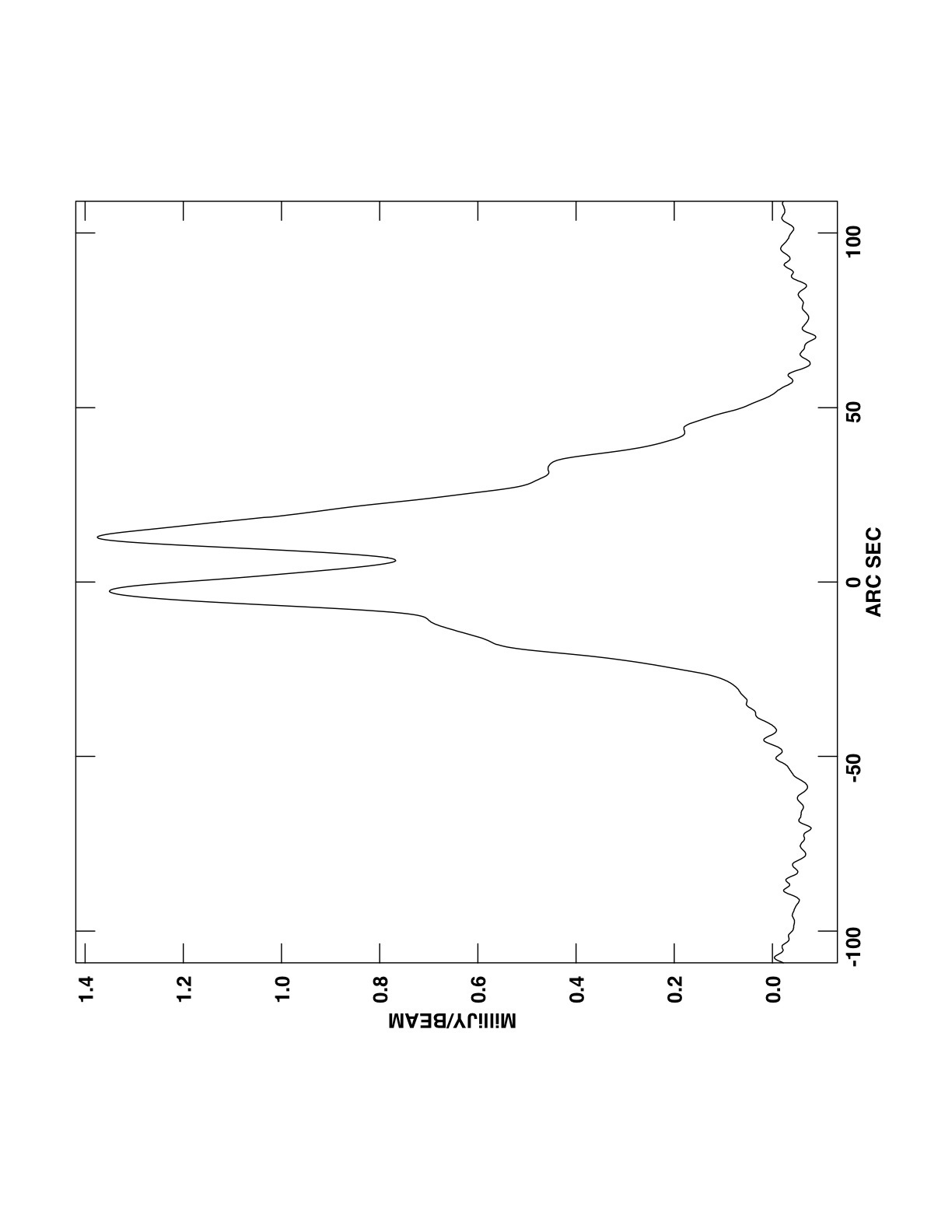}
\caption{VLA 8.5 GHz radio continuum intensity as function of position (in arcseconds) across \g. This slice was taken through the broadest part of the radio ``ring/torus'' on a line that is at a 45\arcdeg~angle (E of N). The beginning point is RA, DEC (J2000): 19 30 34, 18 51 45, and the end point is RA, DEC (J2000): 19 30 26, 18 53 00. This figure illustrates that the central radio emission is concentrated in a ring-like distribution.} 
\end{figure}

\begin{figure}
\includegraphics[width=0.75\textwidth,angle=0]{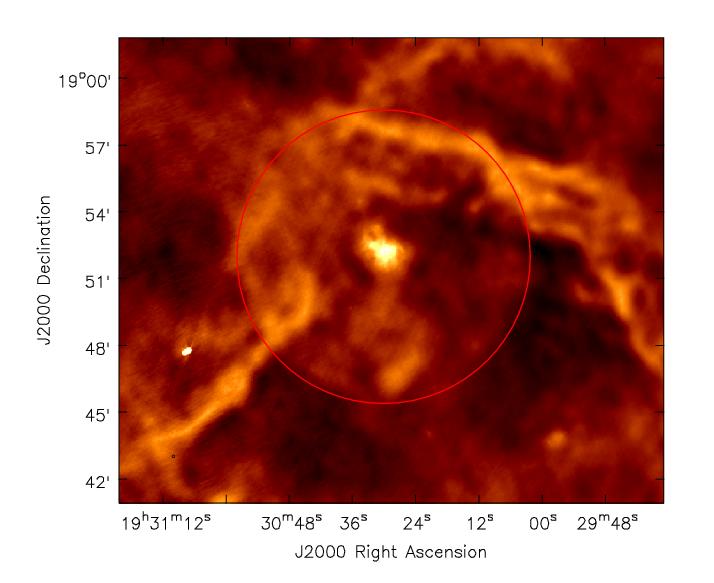}
\caption{VLA 1.4 GHz continuum image of the region surrounding the compact source \g, with a resolution of 6.82\arcsec~$\times$ 6.60\arcsec, PA=5.5\arcdeg. The new radio shell is marked in the figure.}
\end{figure}

\begin{figure}
\includegraphics[width=1.0\textwidth,angle=0]{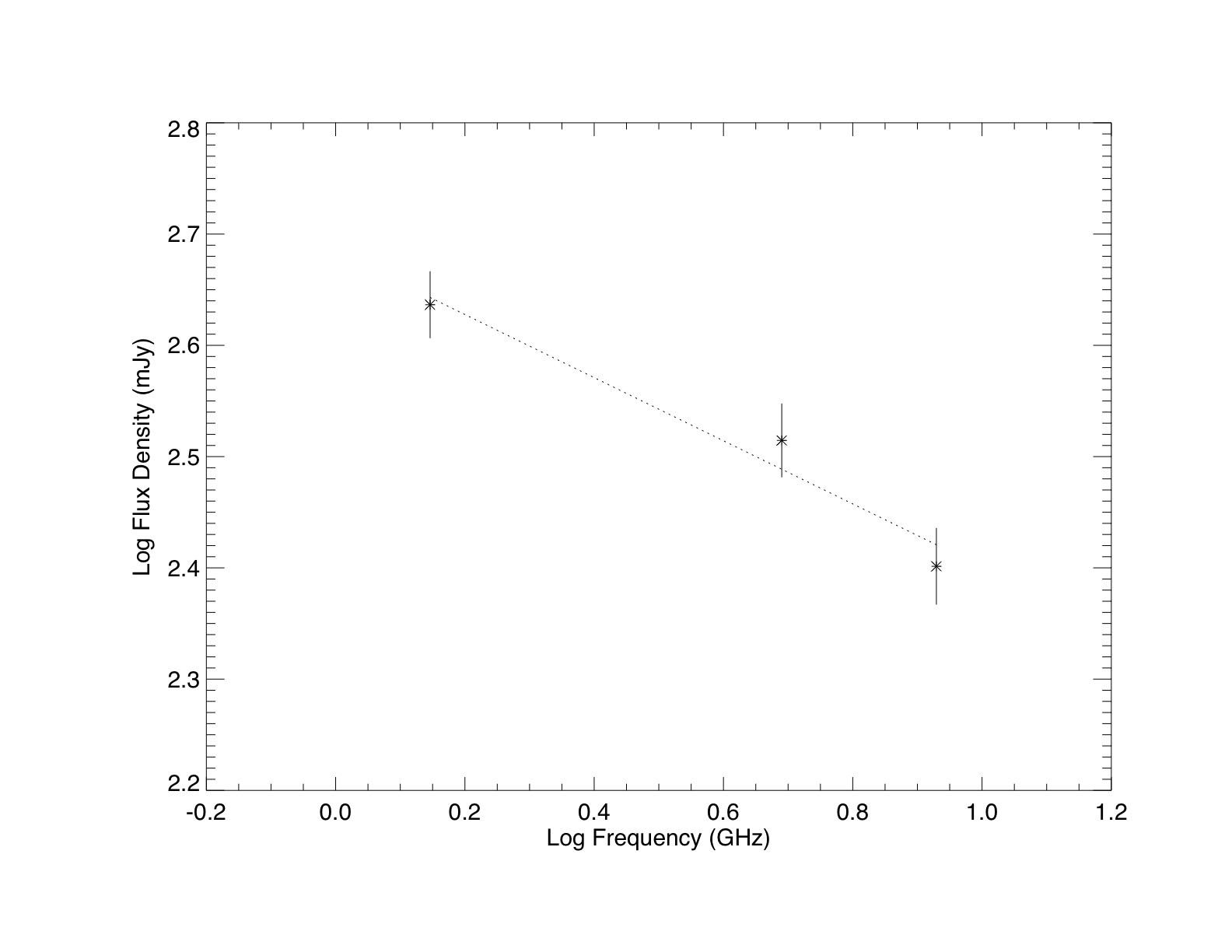}
\caption{Radio spectrum of \g, based on flux densities measured at 1.4, 4.7 and 8.2 GHz and shown with errors. The dotted lines represent the best fit to the spectral index.}
\end{figure}

\begin{figure}
\begin{center}
\includegraphics[width=1.0\textwidth,angle=0]{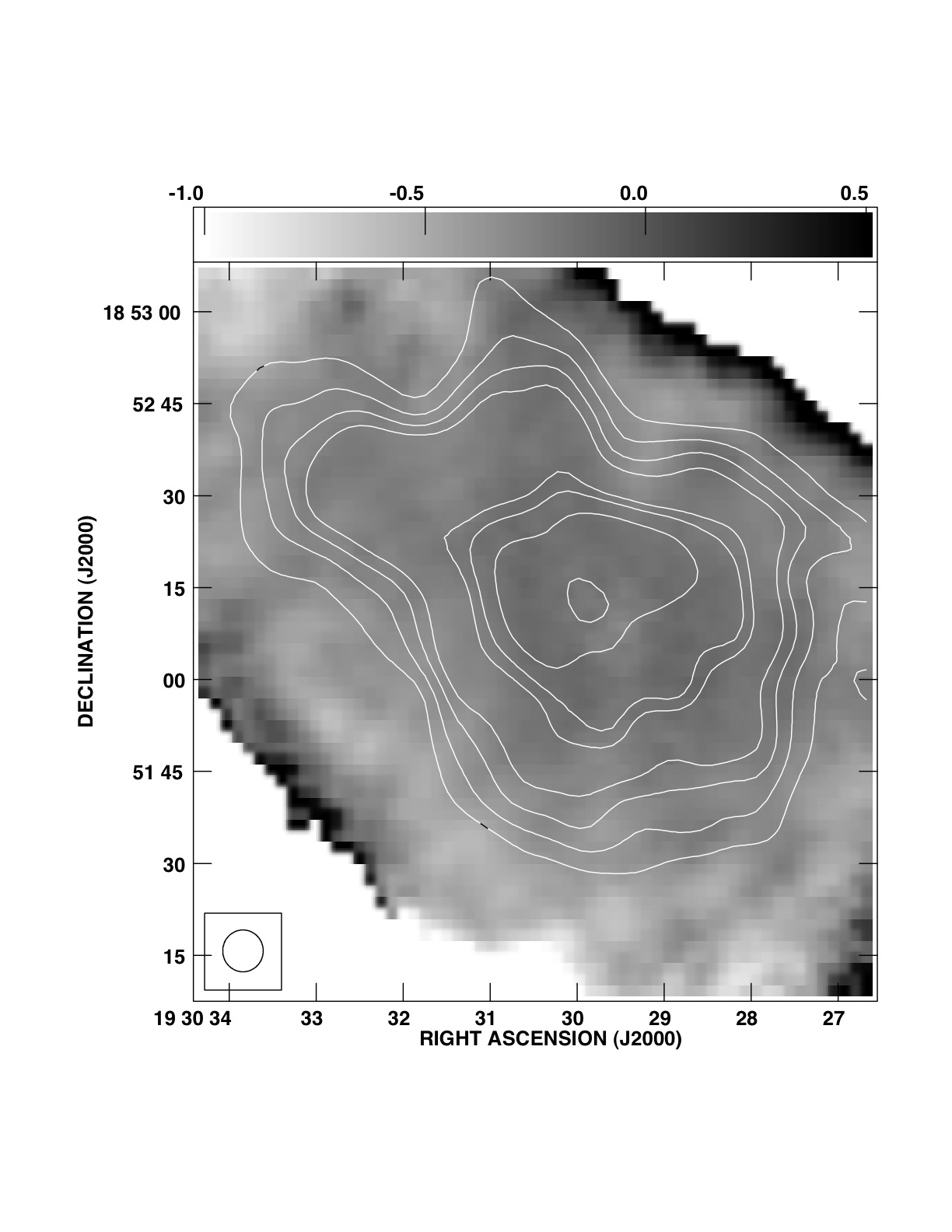}
\caption{VLA 4.7 GHz continuum continuum contours of G54.1+0.3 shown overlaid on greyscale representing 
the 1.4 to 4.7 GHz spectral index, with greyscale that ranges from $\alpha$=+0.5 to $-$1.0.}
\end{center}
\end{figure}

\begin{figure}
\includegraphics[width=0.65\textwidth,angle=0]{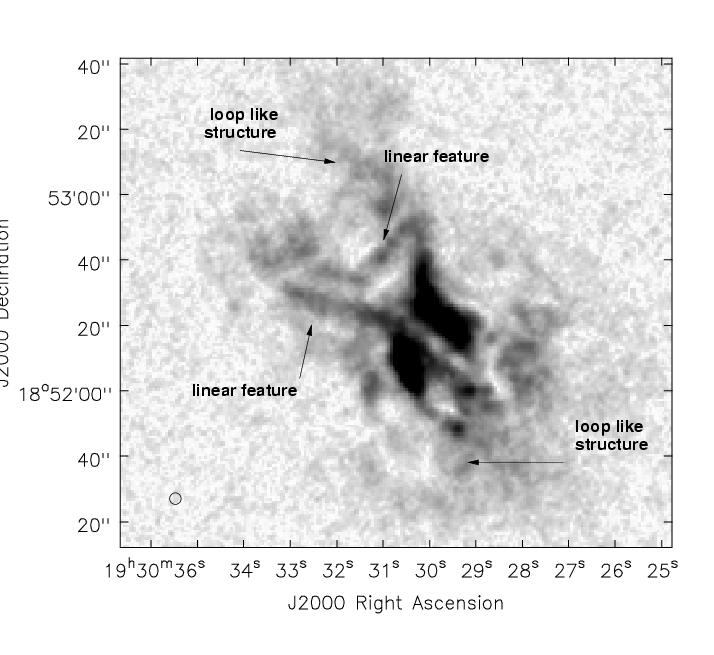}

\includegraphics[width=0.65\textwidth,angle=0]{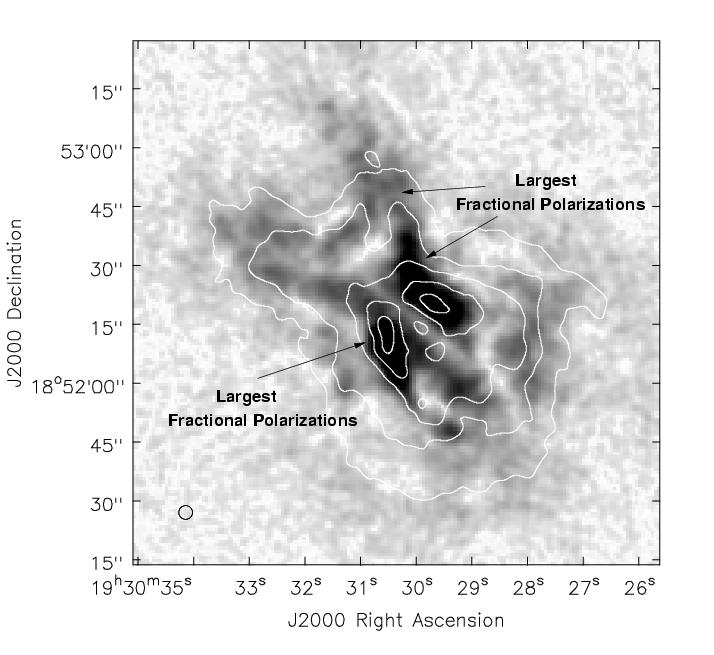}
\caption{Polarized intensity images shown in greyscale, which ranges from 0 to 200 $\mu$Jy \beam, and have resoltions of 3.5\arcsec~$\times$ 3.5\arcsec~at {\bf (top)}: 4.9 GHz and {\bf (bottom):} 8.2 GHz overlaid with contours of total intensity at levels of 0.25, 0.5, 0.75, 1.0, and 1.25 mJy \beam.}
\end{figure}

\begin{figure}
\includegraphics[width=1.0\textwidth,angle=0]{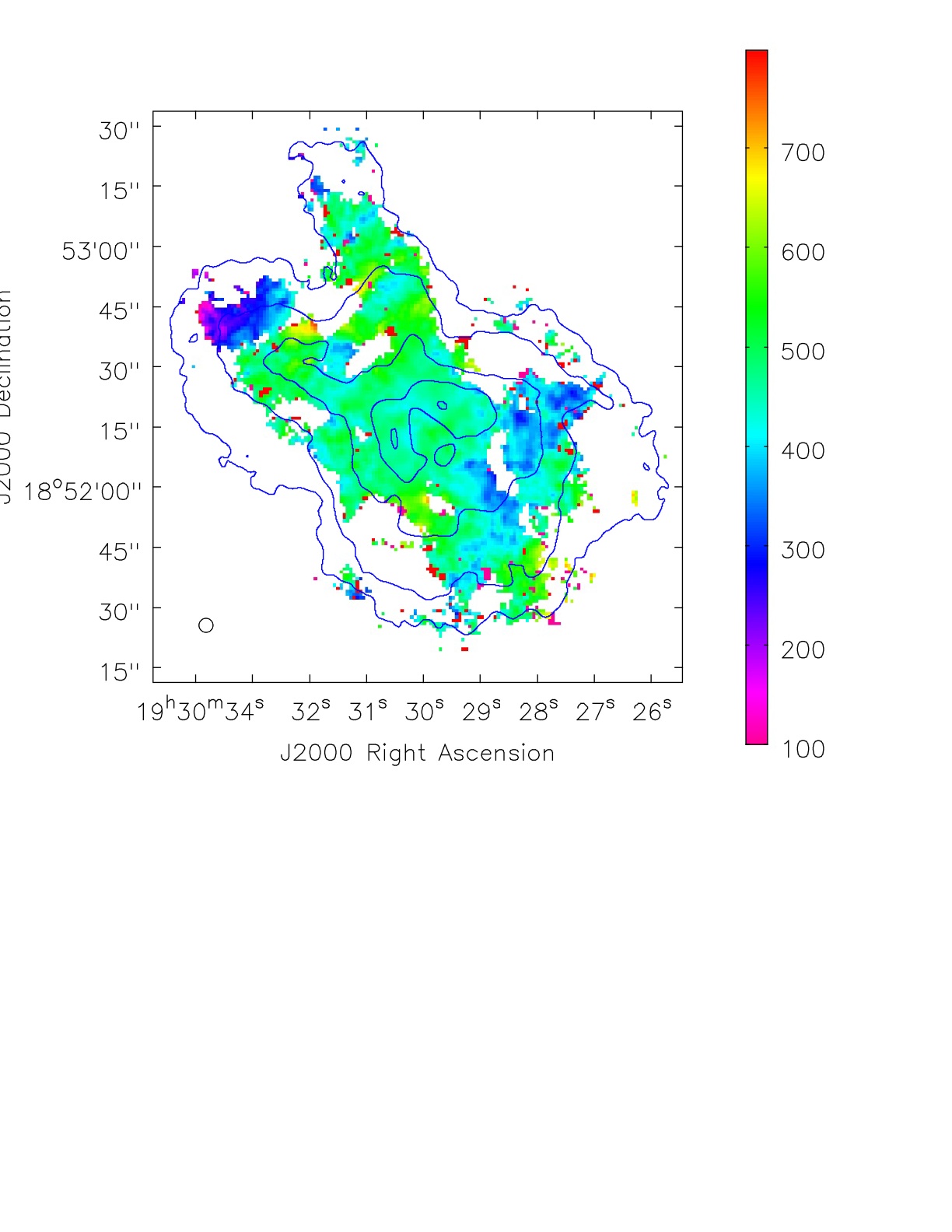}
\caption{Colorscale representing the distribution of rotation measure (RM) toward \g~in units of 100-500 rad m$^{-2}$, with a spatial 
resolution of 3.5\arcsec~$\times$3.5\arcsec.}
\end{figure}

\begin{figure}
\includegraphics[width=1.0\textwidth,angle=0]{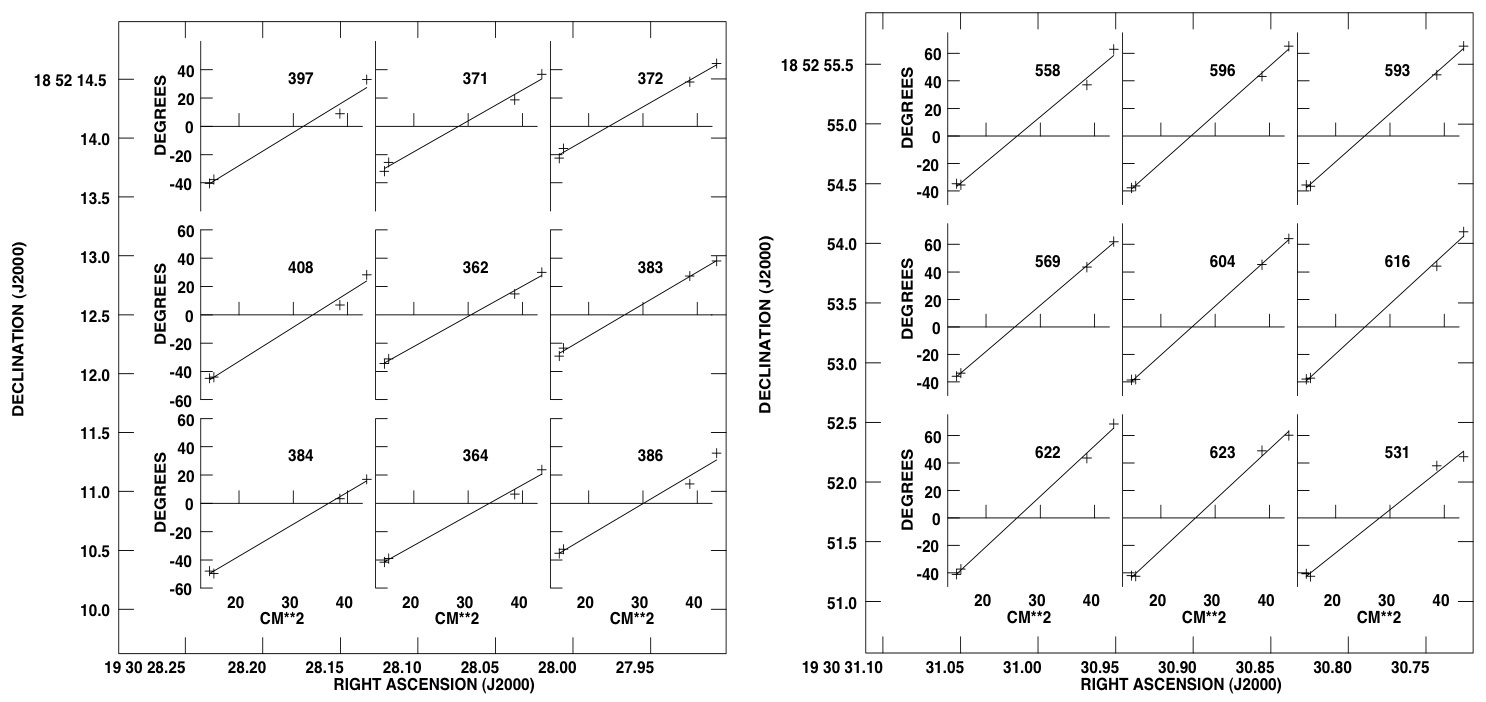}
\caption{Polarization rotation angle versus wavelength squared for two regions in \g.}
\end{figure}

\begin{figure}
\includegraphics[width=1.0\textwidth,angle=0]{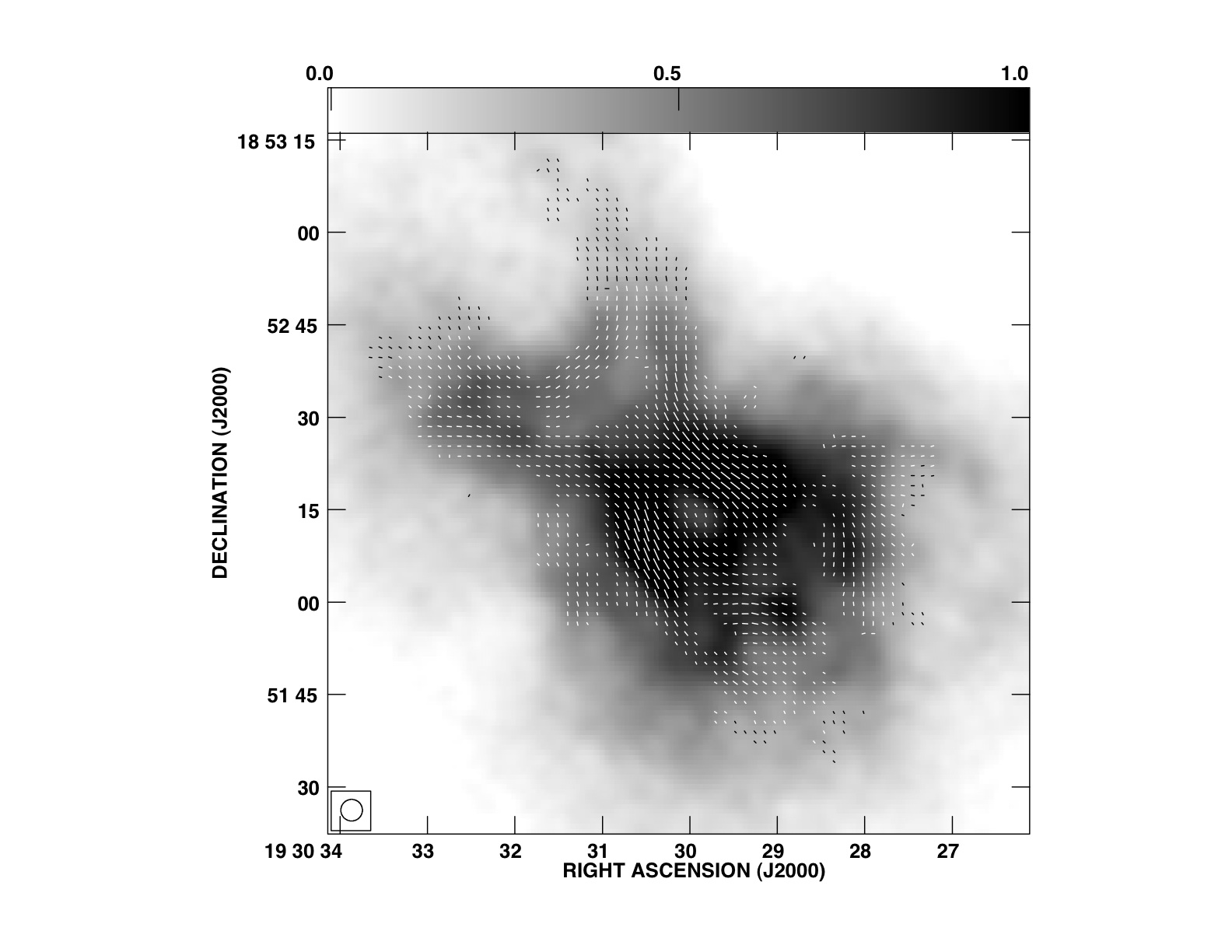}
\caption{Vectors of the intrinsic magnetic field orientation in \g~overlaid on greyscale representing 8.5 GHz emission between 
0 and 1.0 mJy beam$^{-1}$ (see top axis).}
\end{figure}

\begin{figure}
\includegraphics[angle=0,width=0.5\textwidth]{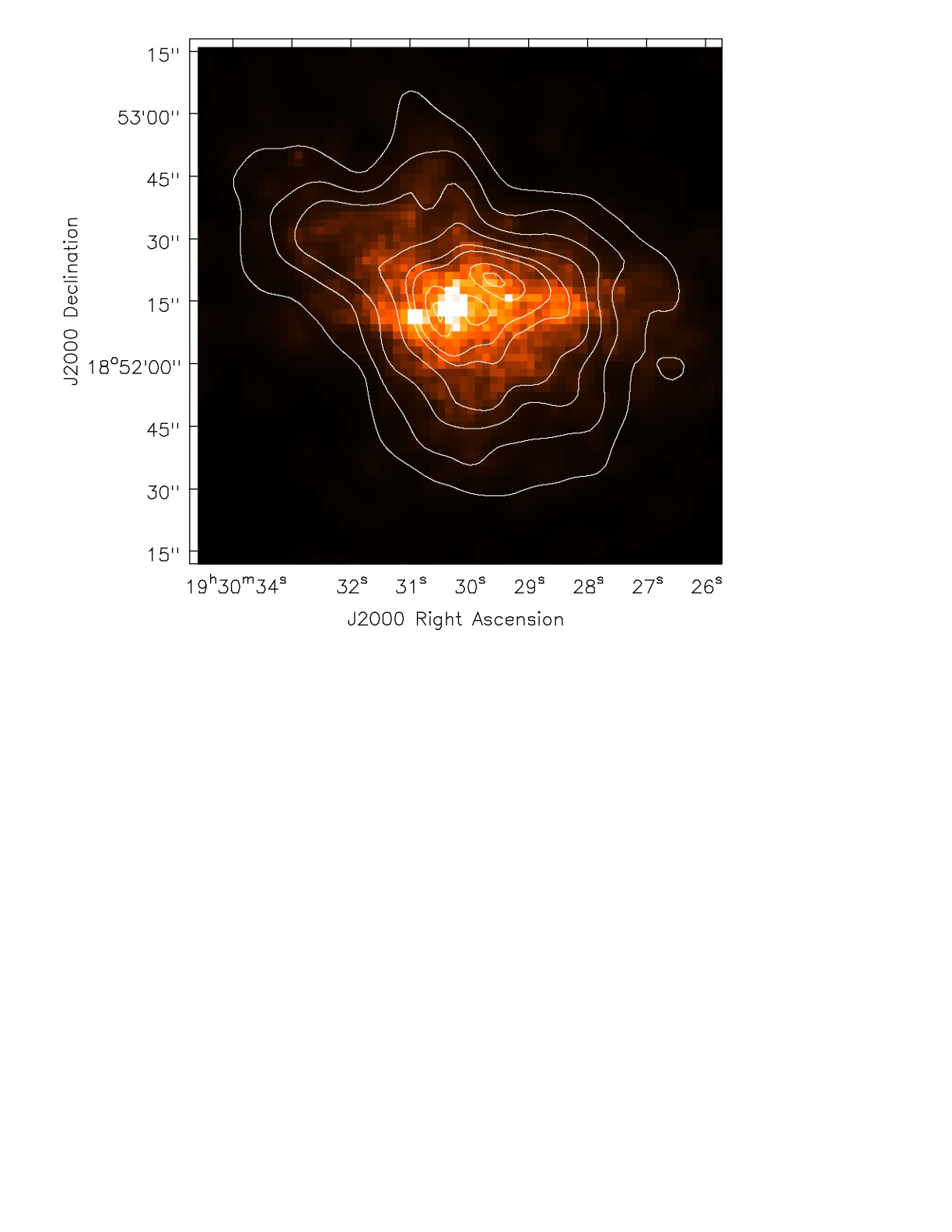}
\includegraphics[angle=0,width=0.5\textwidth]{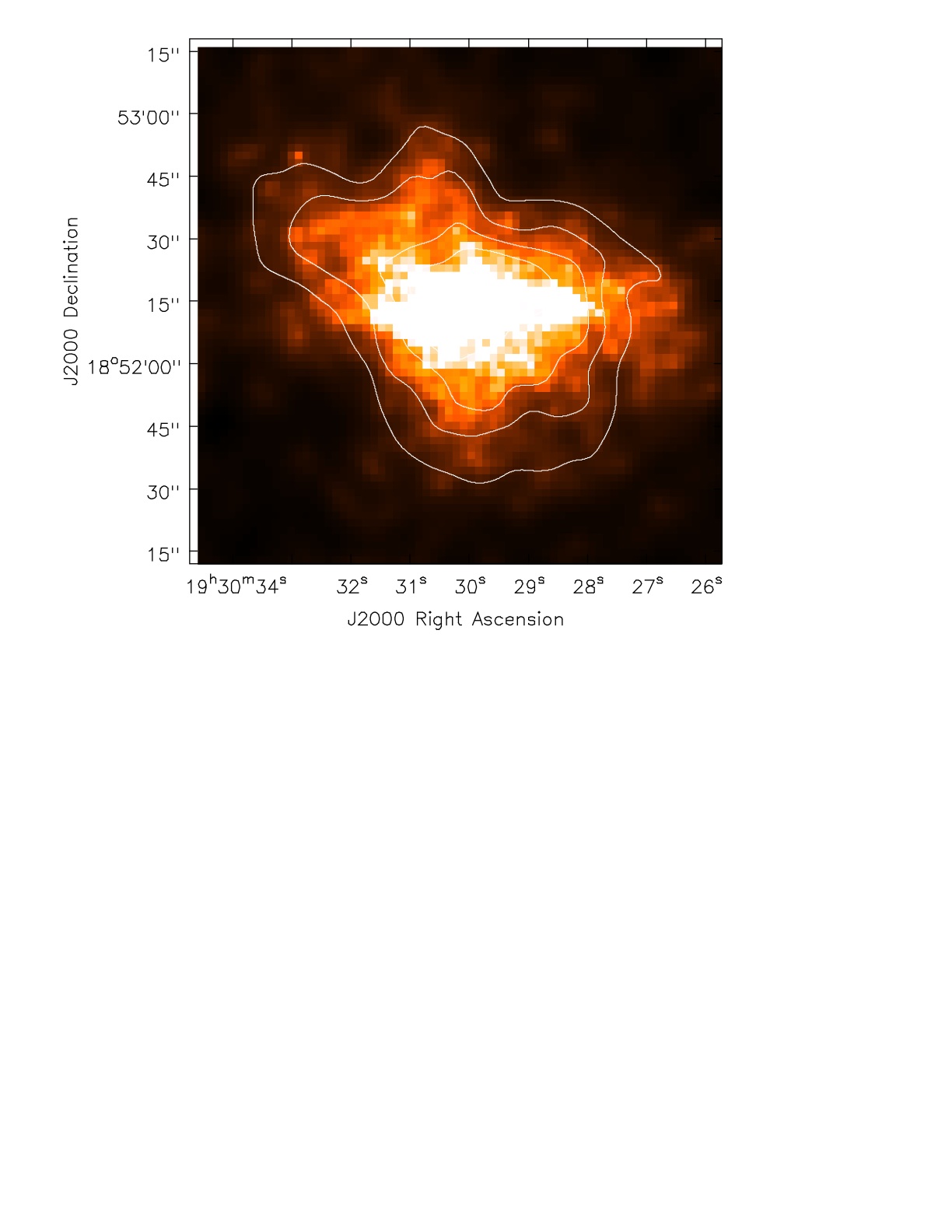}
\caption{Comparison of radio emission (4.7 GHz data from this study; see Figure 1) and X-ray emission from the {\it Chandra} X-ray observatory (Lu et al. 2002). The top image has a color stretch selected to show the centrally-concentrated X-ray emission for comparison with the radio emission in the nebula, and the bottom image has a color stretch set to highlight the diffuse X-ray emission which fills the same volume as the radio nebula.}
\end{figure}

\begin{figure}
\includegraphics[angle=0,width=0.6\textwidth]{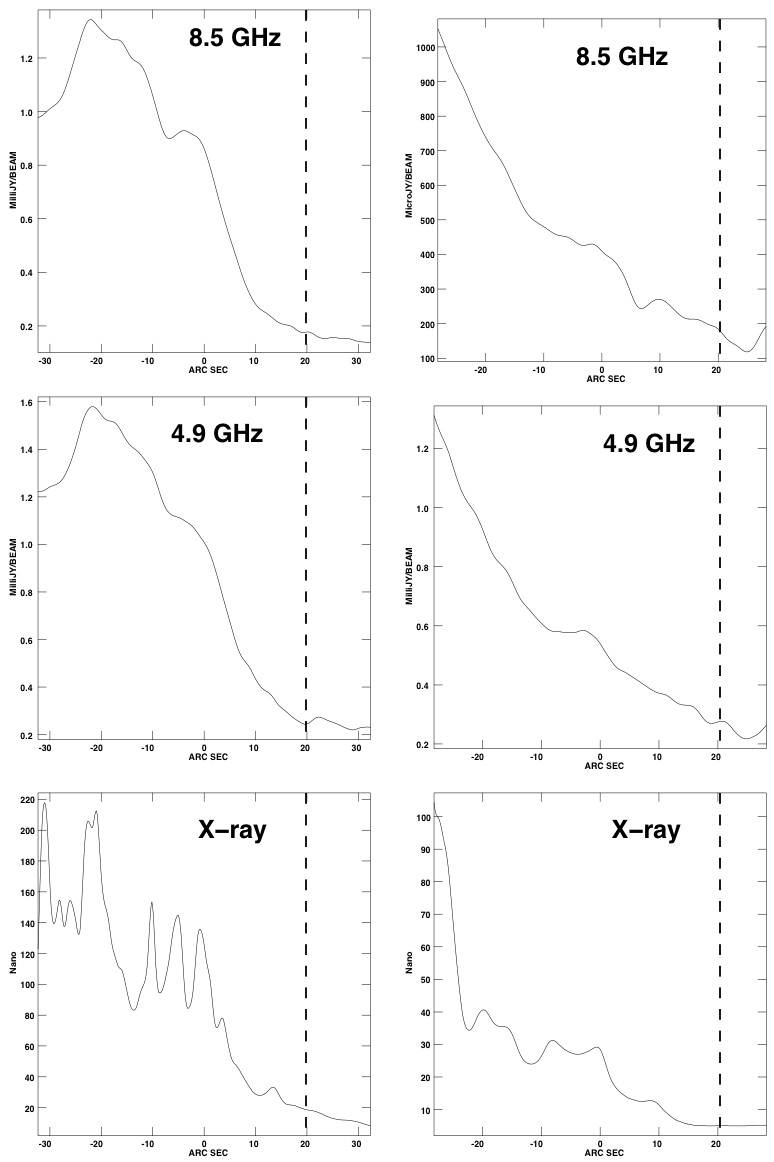}
\caption{Comparison of the radio and X-ray extent of \g. The upper panels are radio frequencies 
(4.7 and 8.5 GHz) and the bottom panels are X-ray emission.  The
brightness profiles show a radial slice from the center of the nebula outwards at two different
locations. The dashed line marks the edge of the radio nebula.}
\end{figure}

\begin{figure}
\includegraphics[angle=0,width=1.0\textwidth]{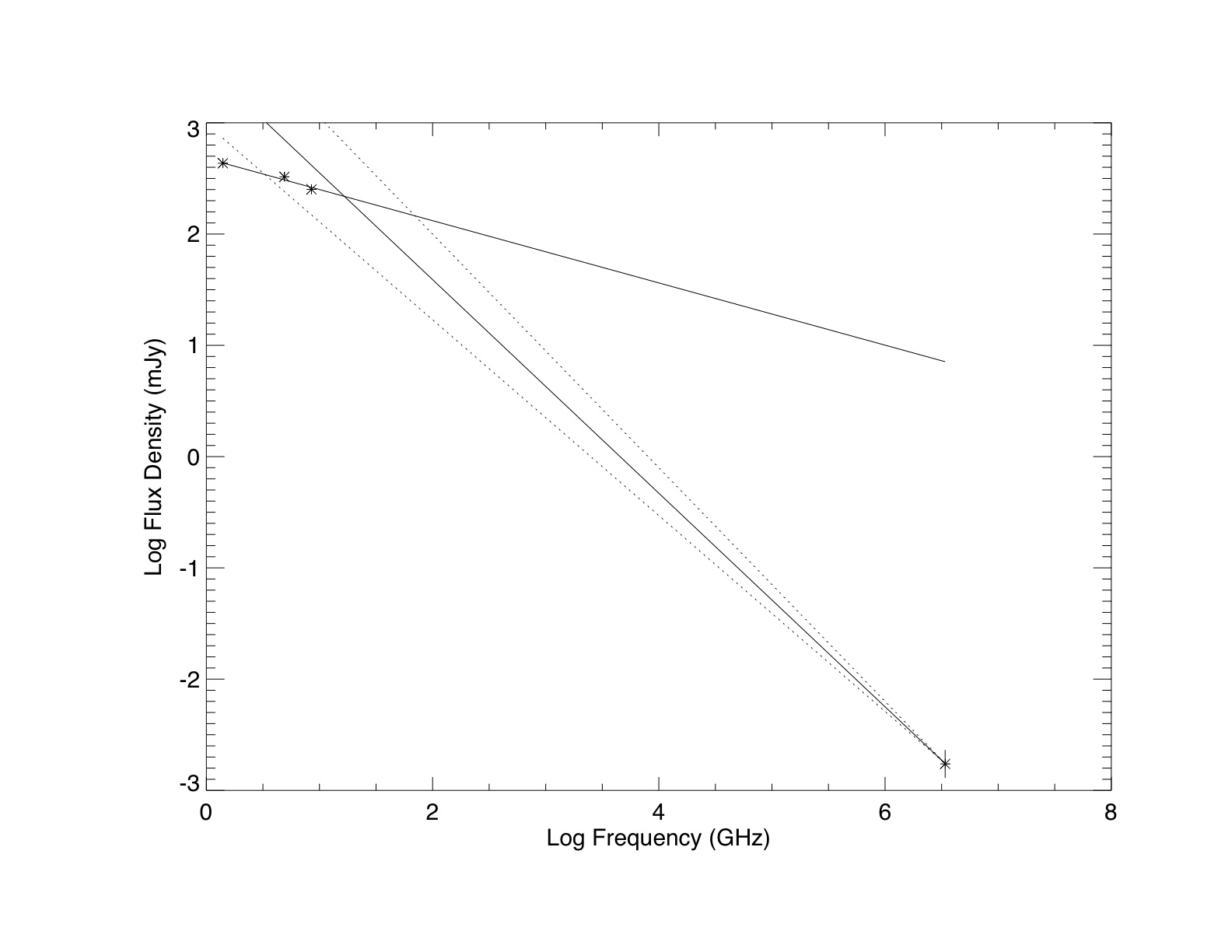}
\caption{Spectrum of the \g~PWN over the radio to X-ray regime, with the radio spectrum ($\alpha_{R}$=$-$0.28)
shown as a solid line, and the X-ray spectrum ($\alpha_{X}$=$-$0.96) shown with a solid line. The position where
these two curves meet is known as the break frequency and occurs near $\nu_{b}$=50 GHz. However, the errors on the 
X-ray spectral fit give a large range of possible break frequencies, from $\nu_{b}$=3 to 200 GHz.}
\end{figure}

\begin{figure}
\includegraphics[angle=0,width=1.0\textwidth]{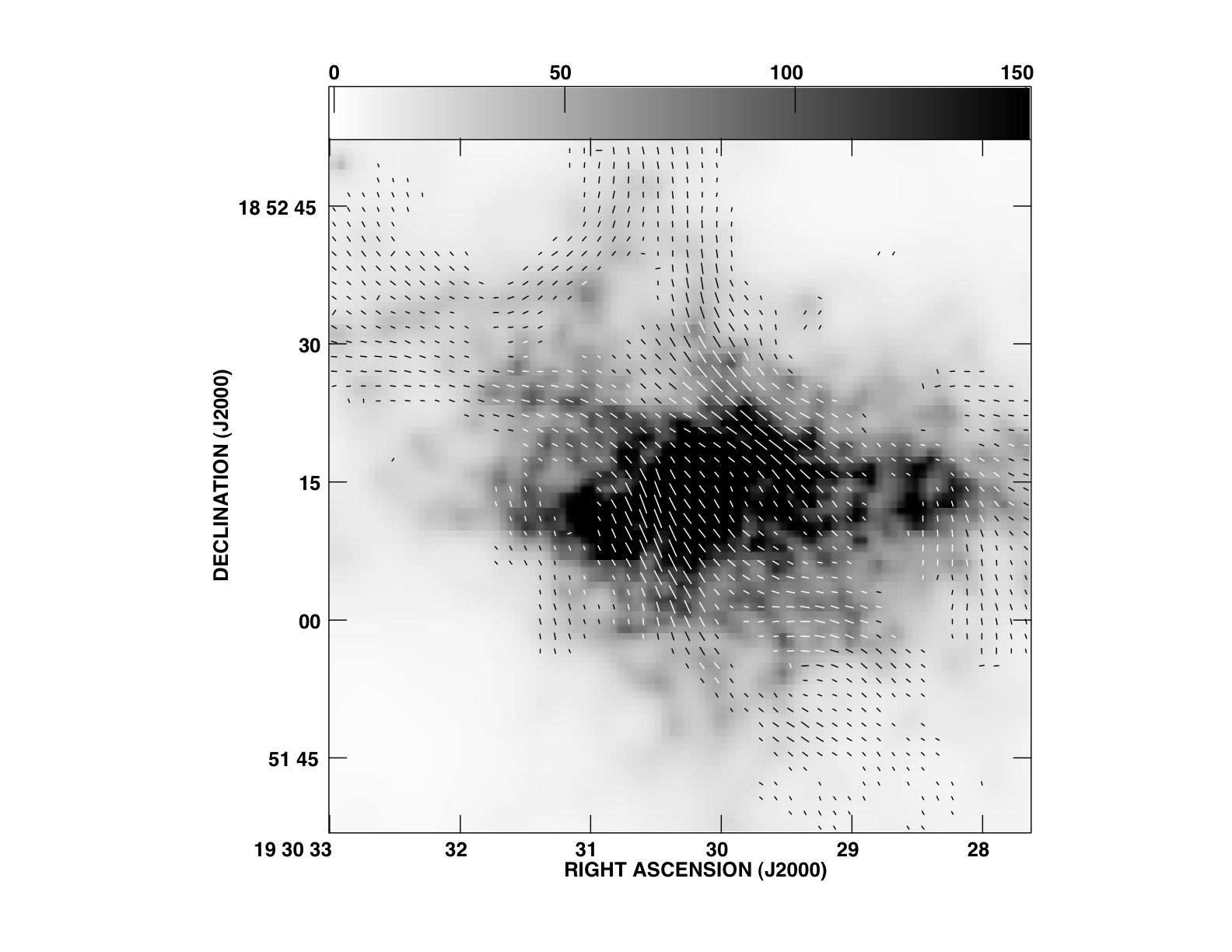}
\caption{Vectors of the intrinsic magnetic field orientation in \g~from these data (presented in Figure 10) 
overlaid on greyscale showing the 0.2-10 keV X-ray emission from Lu et al. (2002).}
\end{figure}

\begin{figure}
\includegraphics[angle=0,width=0.75\textwidth]{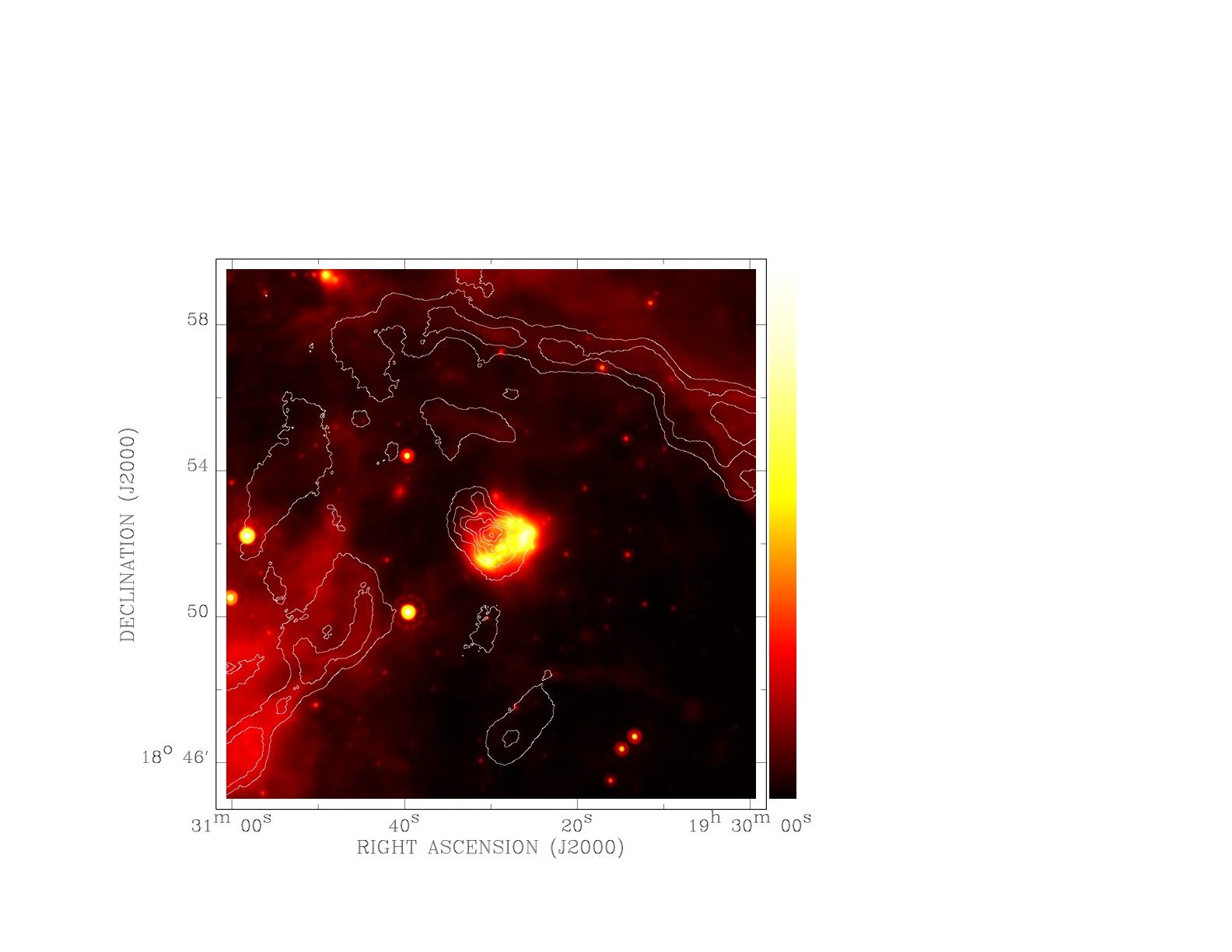}
\caption{Comparison of the large-scale structure surrounding \g~showing 1.4 GHz radio emisson (white contours) and the {\it Spitzer} 24 $\mu$m data from the MIPSGAL survey (Carey et al. 2009). \g~is the compact object in the middle of the image in green contours and the colorscale shows the new shell of radio emission in addition to the local galactic background. This image is shown in Galactic coordinates.}
\end{figure}

\begin{figure}
\includegraphics[angle=0,width=1.0\textwidth]{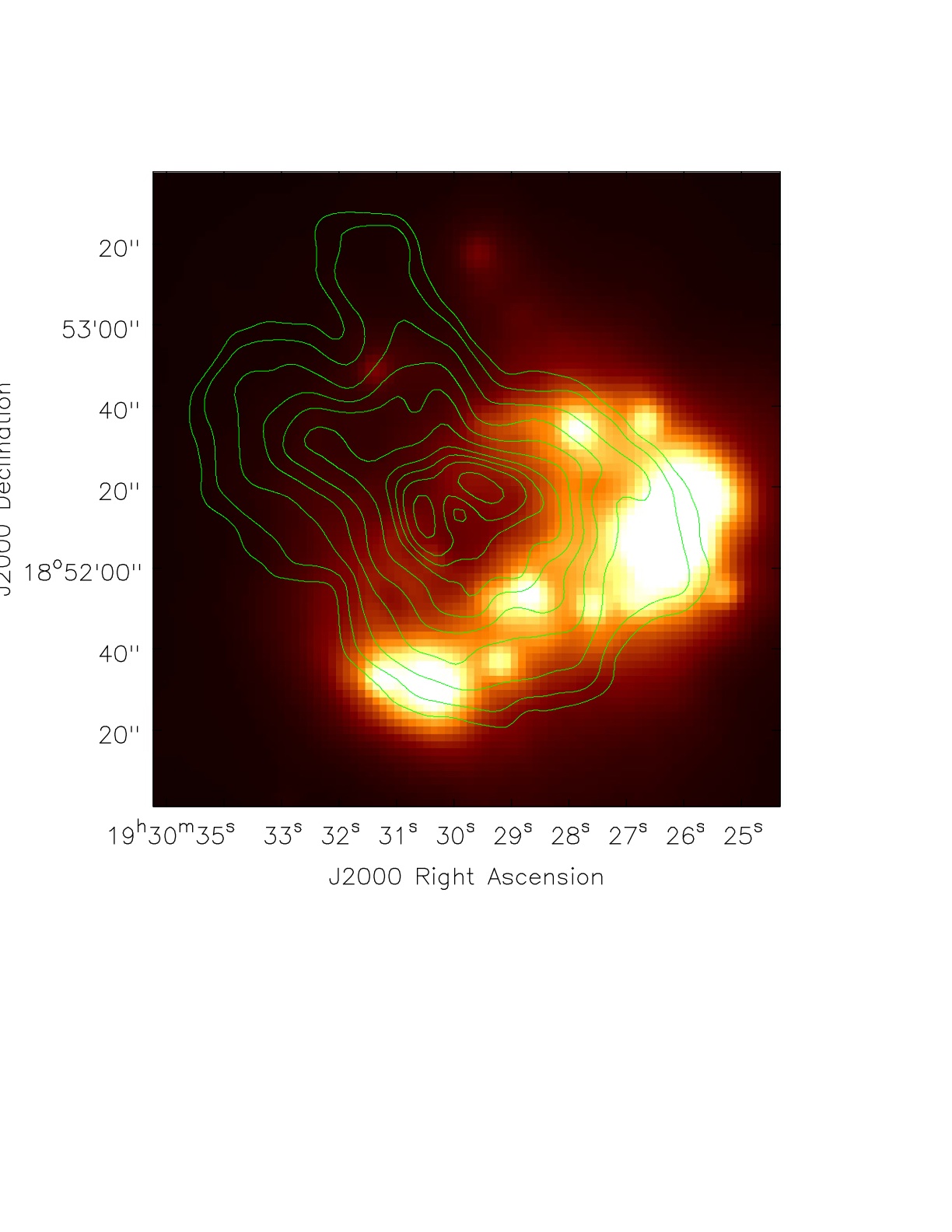}
\caption{Zoomed-in view of the comparison of radio continuum emisson (contours of 4.7 GHz emission) 
and colorscale showing {\it Spitzer} 24 $\mu$m data from the archive (program=3647).}
\end{figure}

\begin{figure}
\includegraphics[angle=0,width=1.0\textwidth]{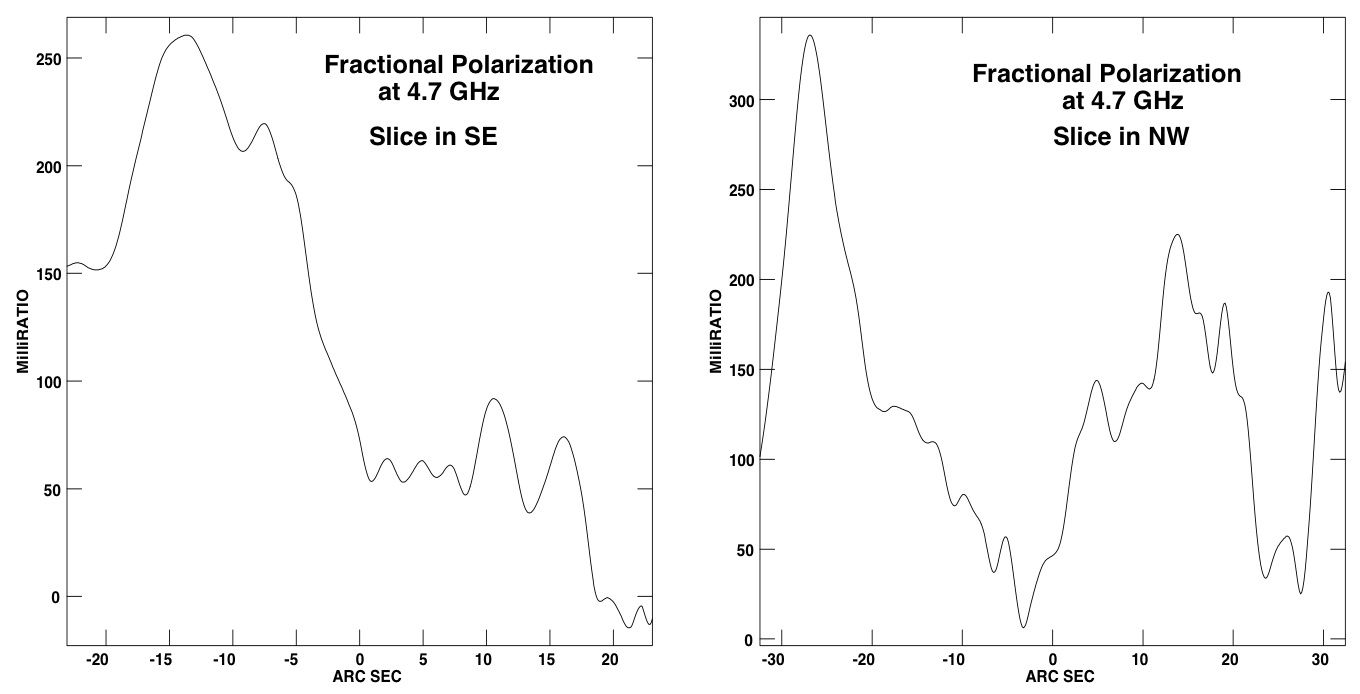}
\caption{Slices in fractional polarization at 4.7 GHz (polarized intensity/total intensity) across \g. (left) Slice
in the SE part of the PWN beginning near the center of the source near RA, DEC (J2000): 18 52 15, 19 30 31, and extending
essentially south to RA, DEC (J2000): 18 52 45, 19 30 30. 
(right) Slice originating at the center of the PWN near RA, DEC (J2000): 18 52 30, 19 30 30 and extending
essentially eastward to the edge of the nebula at RA, DEC (J2000): 18 52 15, 19 30 28. In this slice, it is clear that the
polarization falls off, then increases again before the edge of the PWN.}
\end{figure}

\clearpage
\begin{figure}
\includegraphics[angle=0,width=1.0\textwidth]{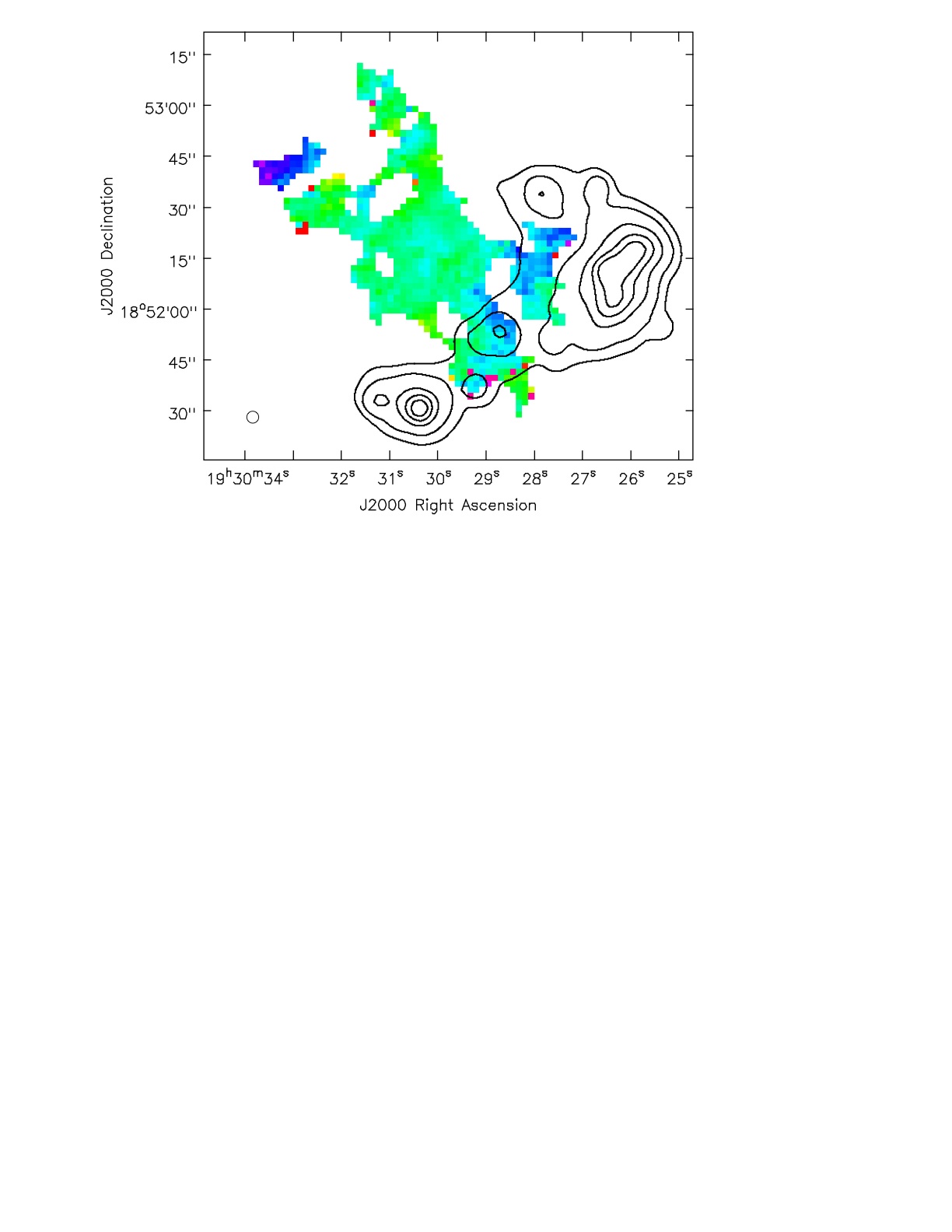}
\caption{Distribution of rotation measure in rad m$^{-2}$ (as shown in Figure 8) with contours 
of 24 $\mu$m emission from the {\it Spitzer} archive (program=3647) overlaid. What is notable here
is the anti-correlation between the rotation measure and the intensity of 24 $\mu$m emission. }
\end{figure}
\clearpage

\begin{figure}
\includegraphics[angle=0,width=1.0\textwidth]{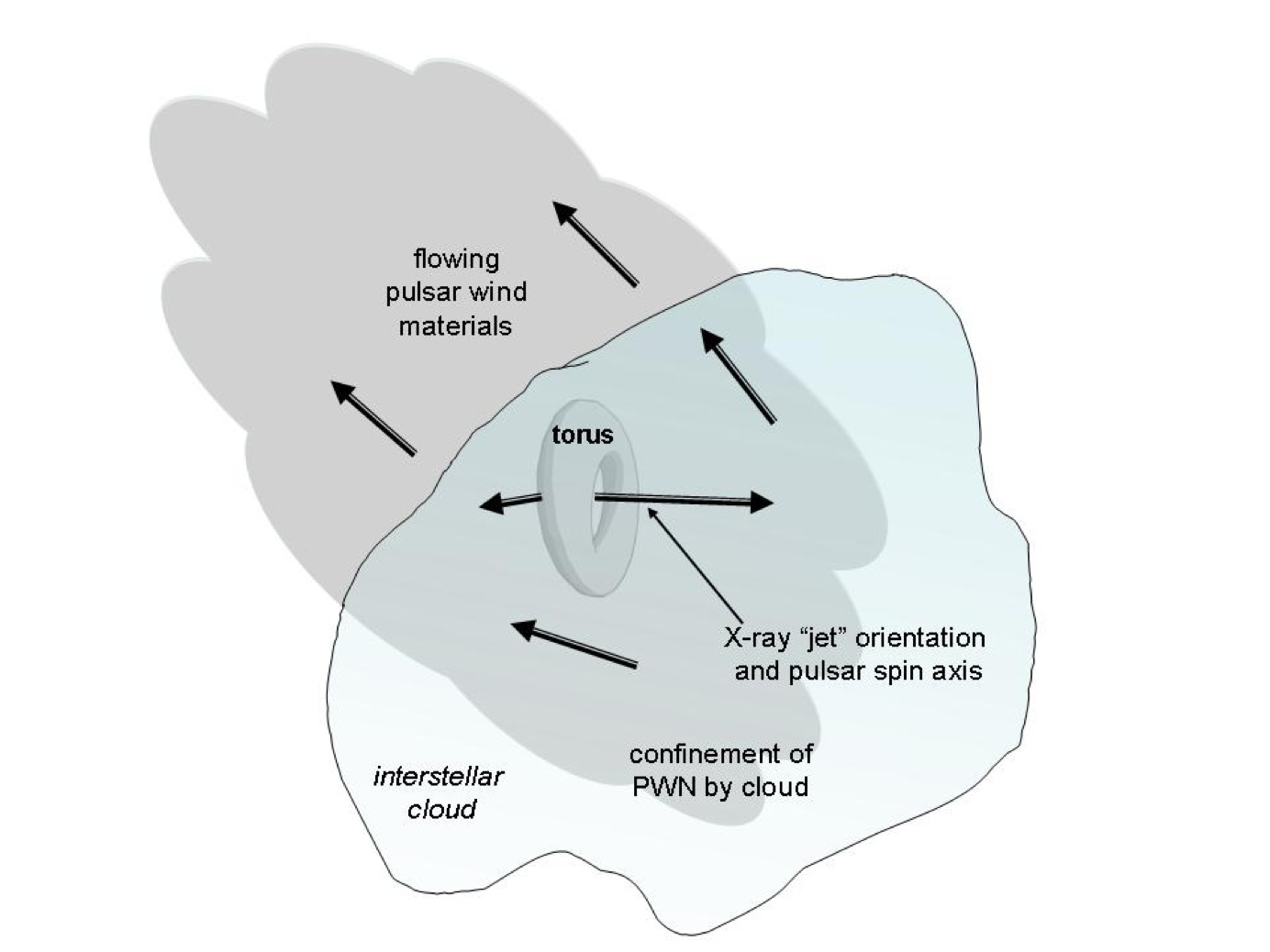}
\caption{Sketch of a possible scenario for the interaction of G54.1+0.3 with an interstellar cloud.}
\end{figure}

\end{document}